\documentclass{article}

\usepackage{arxiv}

\usepackage[utf8]{inputenc} 
\usepackage[T1]{fontenc}    
\usepackage{hyperref}       
\usepackage{url}            
\usepackage{booktabs}       
\usepackage{amsfonts}       
\usepackage{nicefrac}       
\usepackage{microtype}      
\usepackage{lipsum}
\usepackage{graphicx}

\title{Improved Healthcare Access in Low-resource Regions: A Review of Technological Solutions}

\author{
 Bishal Lamichhane \\
  Rice University\\
  Houston, Texas, USA \\
  \texttt{lamichhane.bishal@gmail.com} \\
   \And
 Navaraj Neupane\\
 ASML\\
 Eindhoven, The Netherlands\\
 \texttt{info2navaraj@gmail.com}\\

}

\begin{document}
\maketitle

\begin{abstract}

Technological advancements have led to significant improvements in healthcare for prevention, diagnosis, treatments, and care. While resourceful regions can capitalize on state-of-the-art healthcare technologies, there might be barriers and delays in technology-enabled healthcare availability for a low-resource region. Unique innovations guided by the constraints of low-resource regions are required to truly make healthcare technologies ubiquitous and achieve the goal of \textit{"healthcare for all"}. In this review, we identified several research and development works that have investigated technology-based healthcare innovations targeted at low-resource regions. We found three main pillars of work towards this end: low-cost hardware for the affordability of medical devices, use of information and communication technology (ICT) tools for scalability and operational efficiencies in healthcare services, and mobile health solutions. Several emerging technologies are also promising for healthcare in low-resource regions, such as artificial intelligence,  the Internet of Things (IoT), and blockchain technology. We discuss these emerging technologies too in this review. 

\end{abstract}


\section{Introduction}

The healthcare burden has been rising globally with increased incidences of sickness and poor health~\cite{WHO2018}. We have observed the situation get further dire in the past years due to the global pandemic of COVID-19. While the healthcare needs have only been increasing, the skilled human resources required to meet the associated increasing needs have not been growing proportionally~\cite{Rowe2016}. Technological advancements could help to reduce the burden in the healthcare system, for example by automating diagnosis and improving operational efficiencies. The need to improve diagnosis, treat new diseases and provide care for more people has meant that the state-of-the-art healthcare infrastructure is getting more complex. This increasing complexity has also meant rising costs. The standard healthcare infrastructure of today could be cost-prohibitive in many low-resource regions. Since many people still live in areas that can be categorized as being low-resource regions~\cite{Gupta2015}, efforts toward making healthcare accessible and affordable in low-resource regions should also be a priority. This will help to fully realize the global goals of \textit{healthcare for all} such as the sustainable development goals of good health and well-being~\cite{UN2020}. 

Low-resource regions, also referred broadly to as low and middle-income countries (LMICs), are the geographical regions with a smaller economy and poor access to basic infrastructures~\cite{WHO2012}. While healthcare has several general challenges such as identifying treatments for new diseases, there are some challenges specific to the low-resource settings. Standard healthcare infrastructures being highly cost-prohibitive in low-resource settings is one of such challenges. Even in terms of human resources, the low-resource regions tend to have a smaller health professional to patient ratio~\cite{WHO2018_1}. Skilled healthcare professionals often chose to live in resourceful regions for better facilities and remuneration~\cite{AlShamsi2017}. Hence, it is not uncommon in low-resource regions that personnel without appropriate training for a healthcare role still fill in that role due to the lack of available trained professionals. An example is the non-physician clinicians commonly seen in low-resource regions~\cite{Mullan2007}. This unavailability of adequate skilled health professionals has implications on the quality of healthcare. Besides healthcare professionals, the patients themselves are also a factor in the quality of healthcare delivery. Generally speaking, people living in the low-resource regions might have had lesser access to quality education~\cite{Siddiqi2005}. Thus, they might have less awareness about health, good practices, and seeking care which has implications for the effectiveness of the healthcare that they receive. Similarly, there might be impediments due to cultural and religious barriers which impact health services and delivery in low-resource regions~\cite{Heigener2014}. 

Technology can play a vital role in addressing some of the healthcare challenges in low-resource regions. New devices for diagnosis and treatment could be helpful to streamline healthcare where adequate healthcare professionals are not available. For example, a device for automated respiration rate monitoring could relieve nurses from having to manually estimate respiration rate~\cite{noordam2015use} and enable them to triage more patients in a given time. New technological designs and manufacturing improvements geared for cost reduction can make healthcare devices affordable for low-resource regions. However, there are also open questions regarding the impact of cost reduction on the quality and reliability of the devices (solutions). Information Technology (IT) and automation can also help make healthcare solutions cheaper by scaling existing healthcare solutions and simplifying workflows in healthcare delivery. Similarly, automated clinical decision-making to aid healthcare professionals can increase the efficiency of health processes and reduce associated costs. An important platform for pervasive and affordable healthcare solutions could be the currently ubiquitous smartphones and mobile networks which are getting more accessible in low-resource regions~\cite{Karlsson2018}. Some of the upcoming technologies will also likely contribute to improving healthcare in low-resource regions. For example, advanced machine learning and artificial intelligence have already shown promises for improving and democratizing healthcare~\cite{wahl2018}. Other emerging technologies like the Internet of (Medical) Things (IoT/IoMT) and blockchain are also poised to revolutionize healthcare~\cite{dash2020} and likely lead to the advancement of healthcare in low-resource regions~\cite{upadhyaya2018}.

In this paper, we review some representative works on healthcare technologies for low-resource regions. In Section~\ref{section:low_cost_devices}, we present works on low cost medical devices. Then the use of IT solutions to scale and democratize healthcare resources is presented in Section~\ref{section:ICT_tools}. Smartphone applications geared towards healthcare that could be relevant for healthcare in low-resource regions are presented in Section \ref{section:mobile_health}. In Section~\ref{section:emergin_technologies}, we present some of the emerging technologies that could improve healthcare in low-resource regions. Based upon the reviewed works, we discuss the current state and outlook of healthcare technologies for low-resource regions in Section~\ref{section:discussion}.

\section{Low Cost Medical Devices and Equipment}
\label{section:low_cost_devices}

Medical devices are an important component in the healthcare delivery chain used across diagnosis, treatment, and care. For example, diagnostic devices like vital sign sensors and patient monitors, life-critical support devices like ventilators and pacemakers, among others, are required in general and specialized healthcare services. Effective healthcare delivery requires supporting medical devices. However, the cost of medical devices in a particular care setting can be high. The costs are alarmingly high in higher acuity care settings like ICU (Intensive Care Unit) or OR (operating room) and still relatively high in lower acuity care settings like GW (General Ward). These high costs, both the capital expenditure and operating costs, can in themselves be a big impediment to the wider reach of quality healthcare services in low-resource regions \cite{compton2018}.  

Cost reduction of the medical devices can be achieved by several means such as:
\begin{itemize}
\item Re-design the existing devices to reduce components or use low-cost component alternatives
\item Use alternate modalities to achieve similar functionalities as the standard medical devices
\item Innovative operating model to maximize the utilization of costly devices 
\end{itemize}


In this section, we review some works on low-cost medical device development. The works are categorized based on the approach toward lowering the costs.  

\subsection{Low-cost medical devices with reducing component costs}

The overall cost of electronic components goes down every year \cite{Byrne2017}. The bill of materials for standard medical devices would thus lower too and possibly result in devices being affordable for use in low-resource regions. This, for example, has been presented to be a case by \cite{Guthrie2012} for home-based glucose monitoring devices. Glucose monitoring is essential for monitoring chronic conditions like diabetes, a disease of growing worldwide concern and of increasing concern in low-resource regions too \cite{mohan2020management,cho2018idf,nikpour2021innovative}. According to the authors in \cite{Guthrie2012}, the glucose monitoring devices for home use are getting affordable due to the general trend of lowering prices for semiconductor technologies. Standard electronic components such as microcontrollers, real-time clocks, flash memories, etc. have increased functionality while getting cheaper. This proliferation of semiconductor technologies will further help reduce the overall costs of medical devices and thus barriers to entry of medical devices in low-resource regions.

\subsection{Add-ons to smartphone}

Another approach to reducing the cost of medical devices is to use the ubiquitously available smartphone as an add-on or replacement for some components of the medical devices. The authors in~\cite{Pimentel2014} proposed a low-cost blood pressure measurement device for low-resource settings. The authors suggested the use of smartphones for interfacing with a simple sensory front-end comprising a pressure sensor. With all the signal processing and visualization done in the smartphone, with the power supply too coming from the smartphone itself, the cost of the blood pressure measurement device could be significantly lower than a standard measurement device. Further, the smartphone as an interface helps to provide customized measurement guidance based on a person's medical training, preferred language, etc. Similarly, the authors in \cite{Sinharay2016} presented an approach to using a low-cost device add-on, acoustically coupled to a standard smartphone, for realizing a digital stethoscope. This approach reduced the cost by 150 to 400 times in mass production. The digital stethoscope can be used for various care conditions like the remote assessment of heart sounds from a cardiac patient or elderly patient monitoring at home. A digital stethoscope in the context of the COVID-19 pandemic to diagnose lung pathologies in the low-resource setting has been proposed by the authors in~\cite{jain2021development}. The applicability of digital stethoscopes in various application areas is still being actively investigated in several other works too~\cite{chorba2021deep, ali2021protocol, grooby2022prediction, chen2021toward}.

\subsection{Alternate modalities of measurement}

The price of standard medical devices can also be reduced by developing alternate modalities for diagnosis and treatment, which are comparatively cheaper. The authors in \cite{Lawn2006} developed an intermittent fetal heart rate monitoring device using Doppler ultrasound. It provided a comparative diagnostic value to that of a standard reference device such as the Cardiotocograph, at a much lower cost. The proposed device could operate with alternate power sources such as wind and solar power which made it further suitable for low-resource regions. The authors in~\cite{Cardillo2014} proposed the use of wood-lamp technology, a technology for UV light generation, to detect various skin infections. Early detection of skin infections could be crucial in the diagnosis of medical conditions like HIV/AIDS. The authors outlined various challenges related to technology, business model, etc. for the use of the proposed device. Nonetheless, the proposed diagnosis tool could provide a low-cost (pre-) diagnosis solution, suitable for use in low-resource settings when properly designed for easy use. The authors in~\cite{Haney2017} reviewed various technologies for cancer diagnosis, treatment, and care in low resource settings. The authors found, for example, low-cost and portable commercial ultrasound solutions to be suitable in low-resource settings as a powerful diagnostic imaging tool.  

Low-cost solutions for advanced state-of-the-art technologies have been discussed in several previous works. The authors in \cite{Gouwanda2016} presented a gyroscope-based gait monitoring system that can be employed for use in a non-clinical setting also. Gait monitoring has applications in the detection of the onset of neurodegenerative diseases, among others. IMU (Inertial Measurement Unit) like gyroscopes and accelerometers provide a cheaper alternative to standard references like optical marker-based motion capture technology or force plate sensors. The authors in \cite{Gouwanda2016} indeed found that cheaper alternative methods like gyroscope-based monitoring provided comparable accuracy to other standard reference methods. Similarly, the authors in~\cite{powell2021investigating} found inertial sensors to be a suitable tool to monitor physical functioning tasks in low-resource settings.

A smartphone could also provide an alternative to diagnostic devices for certain medical conditions. The authors in~\cite{Palmius2014} presented a smartphone-based tool for mental health assessment that could be deployed in low resource settings. The developed application uses contextual data like actigraphy and light, along with the available vital sign measurements from third-party devices and validated clinical questionnaires, to assess and track an individual's mental health. As the proposed solution does not require the use of any other diagnostic devices and can be deployed in a scalable way, it is well suited as a pre-diagnostic alternative for use in low-resource settings. Smartphone-based mobile sensing data have been shown to be promising in other mental health applications like schizophrenia~\cite{lamichhane2020patient}, depression~\cite{thati2022novel}, bipolar disorder~\cite{faurholt2021daily}, etc.

\subsection{Commodity hardware and software to improve efficiency}

Technological solutions could also improve workflow and reduce clinicians' burden where the patient to clinician ratio is commonly high. The authors in \cite{Zalzala2011} discussed the Humanitarian Technology Challenge (HTC) on using technology to solve the challenges of healthcare service delivery in low-resource settings. RFID technology is proposed to identify and link different healthcare service interactions, e.g., at a clinic or from a mobile health worker, with a central database of a healthcare resource center. Robust identification of healthcare service interactions and linking to electronic health records is aimed at reducing errors in medical decision-making and enabling proper care planning.

The authors in \cite{Naydenova2015} proposed machine learning techniques for the automatic identification of clinical conditions like pneumonia using heart rate, respiration rate, and other clinical information. Such automatic (pre-)diagnosis will lead to significant workflow improvements and medical error reduction. These will then reduce the costs in healthcare delivery. With the commodification of health devices (e.g., smartwatches consisting of several health sensors), the availability of trained personnel for data interpretation and diagnosis can still be an impediment to effective healthcare services in low-resource settings. Automatic data analysis and (pre-)diagnosis, such as those proposed by \cite{Naydenova2015}, can improve the usefulness of available health data in low-resource settings. Other works such as \cite{Shamiluulu2017} also discussed the use of smartphone applications to support patient diagnosis in low-resource settings, especially where workflow can be burdened by the low doctor-patient ratio. The authors also proposed a machine learning-based tool for automated diagnosis of disease. In a similar approach, the authors in \cite{Mitra2008} presented a mobile telecardiology system suitable for a low-resource setting. In the proposed solution, the ECG strip recorded by an ECG technician is captured as images using a mobile camera and sent to a remote server where it is automatically processed to extract key ECG features. Recommendations, and further referral to a trained doctor if required, are provided based on the ECG parameters and associated automatic assessments of cardiovascular conditions.

In the overall landscape of medical devices, the cost associated with infrastructure maintenance cannot be overlooked. In \cite{Ring2004}, the authors discussed the costs associated with healthcare infrastructure maintenance in a low-resource setting. The infrastructure in this context encompassed not only the medical devices but also the related support structures like the buildings, energy management, waste management support, etc. High costs were found to be mostly associated with the operating costs. The capital expenditure could be less in principle, thanks to the various international aid and support provided to low-resource regions. The authors suggested that the design and planning of infrastructure should be done according to the local circumstances of deployment areas. For example, the availability of components for repair, the possibility of training maintenance personnel, the availability of energy sources like solar plants, and the available monitoring unit for proper aid deployment should be taken into consideration. The authors in \cite{Nabiev2008} discussed some challenges in maintaining computer-based infrastructure in low-resource settings. The authors also note that the cost of required trained professionals to sustain and maintain such infrastructure is the most significant. The authors proposed a centralized technological support model for cost-effectiveness that could be relevant for healthcare infrastructure deployments in low-resource regions.

\subsection{Considerations for low-cost devices}

Multiple issues need to be taken care of in the effort toward affordable medical devices for low-resource settings. Proper considerations of the underlying technology (suitability/reliability), business and cost models (sustainability/affordability), associated quality and usage factors, etc. need to be made. The authors in \cite{Cordero2014} advocated the need to approach medical device development for low-resource settings differently from a mere technology transfer process. A sustainable approach to the introduction of new technology is suggested by considering different factors such as customer needs in low-resource regions, the context of device/application usage, viable business models, etc. The authors in \cite{Neighbour2014} raised various issues that could arise in low-cost medical device development. The authors suggested that many of the proposed low-cost devices could be bogus and lack adequate safety designs. Therefore, proper scrutiny of propositions to ensure patient safety and quality care is recommended. The authors also remarked that actual device development costs might not be properly accounted for in many low-cost medical device propositions. Many low-cost medical device development initiatives are undertaken in the academic research setting, usually supported by an external grant. Thus, the research expenditure for a new low-cost device might not always be transparent and fully accounted for. This has implications when new device development projects in low-resource regions are planned. 

\subsection{Section summary}

To summarize, several works have investigated low-cost medical device/solution development for low-resource regions. Some works have relied on the decreasing cost of semiconductor technology to argue that medical devices are getting affordable for use in low-resource regions. This is true for many consumer-facing healthcare technologies (e.g., smartwatches and home-based health monitoring devices). However, much complex medical equipment is still very costly~\cite{vasan2020}, and it might be a while before the prices for standard medical devices required in a health care center become affordable in most low-resource regions. Other works have relied on a smartphone as an add-on for medical devices, overcoming the need to procure standard devices (e.g., a smartphone-based digital stethoscope in place of a standard stethoscope). Though this approach is a viable alternative, e.g., using a smartphone display as a  medical device display, the functionality of the proposed solutions must be rigorously validated. For example, it should be ensured that the functionality of the proposed solution/device works across different smartphone types, network quality, etc. Some works have investigated alternate modalities for affordable diagnostic and therapeutic solutions in low-resource settings. These research works should continue as low-cost measurement modalities might replace costlier alternatives. Other advances we are making in data science and analytics for robust measurements could be helpful. For example, thanks to the advances in data science, photoplethysmography (PPG) from consumer devices can potentially be used for arrhythmia detection \cite{paradkar2017,chang2022atrial} and consumer cameras can be used as a vital sign monitoring device \cite{antink2019,bae2022prospective}. Technological solutions have also been proposed to improve healthcare workflows and improve efficiency. This reduces the overall healthcare costs and thus needs further exploration of their applicability in low-resource settings. Finally, a holistic view of the cost of medical device deployment considers the cost of infrastructure in low-resource settings. This has been discussed in some previous works and must be a part of any proposed general healthcare solution targeted at low-resource regions. Overall, no matter what approach (or a combination thereof) is used towards low-cost medical devices, one must ensure that the quality and patient safety aspects are kept as the priority.

A summary of our observations from the literature regarding low-cost medical devices for low resource regions is provided in Table~ \ref{table:low_cost_devices}. 

\begin{table}

\caption{Summary of works on low-cost medical devices for applications in low-resource settings. Various approaches have been proposed in the literature for reducing the costs, ranging from using alternative sensing technology to the usage of a smartphone as an add-on for medical devices.}
\label{table:low_cost_devices}

\centering
\begin{tabular}{|p{7cm}|p{7cm}|}
\hline 
\textbf{Approach for low cost medical devices and healthcare solutions} & \textbf{Example Clinical Applications and Example Works} \\ 
\hline 
Reliance on reducing the price of the semiconductor technologies for  & Glucose monitoring at home for Diabetes care. e.g., \cite{Guthrie2012}. \\ 
\hline 
Smartphones as an add-on for medical devices & Blood pressure monitoring, Remote cardiac monitoring, and Elderly patient monitoring. e.g., \cite{Pimentel2014},  \cite{Sinharay2016}.
 \\ 
\hline 
Low cost alternative for diagnosis and therapy e.g. using alternate signal for diagnosis & Fetal Heart rate monitoring, Skin infection diagnosis, Cancer diagnosis, Gait monitoring, Mental health assessment. e.g., \cite{Lawn2006}, \cite{Cardillo2014}, \cite{Haney2017}, \cite{Gouwanda2016}, \cite{Palmius2014}
 \\ 
\hline 
Improvement of workflow to bring cost reductions and reduce burden on clinicians & Automated Electronic health record, Automated Diagnosis, Mobile telecardiography. e.g.,  \cite{Zalzala2011}, \cite{Naydenova2015}, \cite{Shamiluulu2017}, \cite{Mitra2008}
 \\ 
\hline 
\end{tabular} 

\end{table}

\section{ICT Tools to Scale Healthcare Solutions}
\label{section:ICT_tools}

The use of information and communication technologies (ICT) in health services has been termed eHealth by the World Health Organization (WHO)~\cite{WHO_ehealth}. Cost-effectiveness and security are important aspects of eHealth propositions~\cite{Iluyemi2008}. Healthcare outreach, especially in low-resource regions, could be improved with ICT to scale and democratize care. For example, ICT could help the rapid dissemination of clinically relevant information such as a new diagnostic procedure or matters of public health. In this section, we review some works that have proposed the use of ICT tools for affordable healthcare in low-resource settings.  

\subsection{ICT for Electronic Health Records (EHR)}

Proper management of health-related data is crucial in overall healthcare delivery. Electronic health records (EHR) and other healthcare information systems provide the required infrastructure for data management and information flow. However, proprietary healthcare information systems might be cost-prohibitive for low-resource regions; the burden of software license costs can sometimes be very high. The proliferation of affordable ICT solutions with the ongoing open-source movement could prove a boon for healthcare ICT infrastructure in rural areas. Several works have identified how ICT advancements can be used for better data management and information flow for healthcare services in low-resource settings.  

In~\cite{Omary2009}, the authors discussed the use of free and open-source software like Care2x~\cite{care2x.org} and OpenEHR~\cite{openehr} to build hospital information systems, in the context of healthcare services for low-resource regions in Tanzania. The authors in~\cite{Shah2013} proposed an open-source health information system called \textit{DataPall} for rural healthcare solutions in Africa. The \textit{DataPall} system for keeping patient care records could run on a PC without internet access. The authors highlighted the role of healthcare information systems such as \textit{DataPall} in providing evidence-based healthcare improvements. The authors in~\cite{Lakshmi2011} also proposed an open-source cloud-based system named \textit{Eucalyptus} to assist healthcare in low-resource regions. \textit{Eucalyptus} could support both the information management needs as well as the elastic computational needs. The information management solution within \textit{Eucalyptus}, for example, is obtained by integrating an open-source hospital management system such as Care2x. The reliance of low-resource regions on open-source solutions for EHR system implementations was also discussed in~\cite{Torre2013}. In particular, the following open-source EHR/EMR (Electronic Medical Records) systems were reported to be commonly used: HOSxp~\cite{hosxp}, OpenEMR~\cite{openemr}, and OpenVista~\cite{openvista}. The authors in~\cite{berrueta2021maternal} identified 8 different data collection systems used for maternal and neonatal care in the low-resource region of which four were free and open-source systems, highlighting the importance and prevalence of open-source healthcare data management solutions in low-resource settings. The authors in~\cite{purkayastha2019comparison} qualitatively compared five popular open EHR systems: OSHERA VistA, GNU Health, OpenMRS, OpenEMR, and OpenEHR. The OpenEMR was found to be the most promising open-source EMR solution in comparison. It is generally beneficial to have multiple open-source projects for an application. A hospital or clinic in low-resource settings could choose from multiple options of open-source EMR/EHR depending upon the fit to their current needs. In the meantime, open-source EHR should be continually improved to be user-friendly, reliable, and effective, given the increasing usage of these EHRs for critical healthcare services in low-resource regions, among other places.

It is preferable if most of the (software) components proposed in a healthcare information system, e.g., within a proposed EHR, are open-source. For example, the authors in~\cite{Goel2017} proposed \textit{Intelehealth} system for telehealth access in rural India. \textit{Intelehealth} is an open-source platform enabled by cloud-based electronic health records. The platform allows proper information flow between different healthcare professionals and clinical centers. MySQL database, available at no cost, is used as a component of \textit{Intelehealth} which makes the proposition suitable for low-resource regions. In~\cite{Sharwardy2013}, the authors discussed a teleconsultation system designed for low-resource regions of Bangladesh which allows the physicians to remotely access patients' records and provide appropriate consultations. The system was built using PHP and MYSQL server components which are both freely available IT tools. Similarly, in~\cite{Pambudi2004}, the authors proposed using open-source connectivity and database system for connecting different healthcare centers and professionals in low-resource regions of Indonesia. Internet radio packet was used for connectivity and PostgresSQL for database management. The free/low-cost software components make the proposition suitable for use in low-resource settings. On the other hand, the authors in~\cite{Chen2011} proposed to use an off-the-shelf database like \textit{Filemaker}~\cite{filemaker} to build an EHR/EMR solution for low-resource settings. However, the \textit{Filemaker} solution is not free and the added cost of the database has to be assessed in the trade-off of costs versus better functionality or usage benefits that a commercial database could provide.  


\subsection{ICT to improve secondary healthcare functions}

ICT tools could help in making secondary healthcare functions like knowledge management, information dissemination, and policymaking affordable and accessible. These secondary functions are especially important for healthcare operations in low-resource regions where the availability of trained personnel is limited. Several studies have discussed the role of ICT in knowledge management, information dissemination, and policymaking in the context of healthcare. Similarly, the role of ICT to improve logistics in healthcare has also been highlighted by earlier studies. Improved logistics in healthcare leads to improved accessibility.

The authors in~\cite{Manya2018} discussed how health information tools have benefited healthcare services in low-resource regions of Kenya. The improved data availability for local bodies helped to better assess the current healthcare systems, identify their bottlenecks, and propose improvements. ICT tools helping improve policy by monitoring the healthcare system are highlighted. In~\cite{Nunziata2002}, the authors discussed the role played by ICT in the maintenance and operations of healthcare infrastructure deployments in Mozambique. As these examples depict, information assimilation enabled by ICT helps in policy and decision-making at a regional/national level. The authors in~\cite{Martinez2005} also reported how ICT is important for improved healthcare outreach and efficiency in low-resource regions of Peru and Nicaragua. They also identified the need for increased access to communication infrastructure, information sharing, and continuous training of health professionals; ICT will have an important role in fulfilling these needs too. Interesting examples are shared by the authors on how ICT could relieve the burden on healthcare professionals. A healthcare center head, who otherwise would have needed to travel and thus created delays in the health services, could avoid the travel when facilities for emails and communications are easily available in a low-resource region. Similarly, many healthcare professionals could benefit from ICT-enabled solutions that reduce errors arising from manual paper works in healthcare. In~\cite{Chaamwe2010}, the authors also discussed the role of telehealth systems built using (open-source) ICT tools to improve healthcare in low-resource regions of Zambia. The proposed solution provided improved access to medical specialists, increased opportunities for education, and logistic improvements like decreased travel needs for both the healthcare professionals and patients. Though the examples discussed in this section are sometimes from a work of more than a decade ago, some analogous technology-based assistance for improved logistics would still exist today.  

\subsection{Network and server resources for healthcare solutions}

The increasing availability of ICT hardware, services, and infrastructures like telecommunication is a prerequisite for improved healthcare availability in low-resource regions. Several previous works have reported how the availability of servers and telecommunications is improving e-health and telehealth services for low-resource regions. The authors in \cite{Iluyemi2008} discussed the role of telecommunication infrastructure and internet access devices in the context of health services in Africa. The authors highlighted how a basic network infrastructure enabled by telecommunication technologies like GPRS, WiMAX, and VSAT provides a backbone on top of which healthcare services and information delivery can be deployed. ICT's role was identified to be crucial in various areas such as health system performance monitoring, resource planning, dissemination of health knowledge, policy-making, etc. The authors in~\cite{Grainger2006} discussed the issue of communication infrastructure affordability. A combination of a Wi-Fi and WiMAX network was proposed as an affordable network solution for low-resource regions. Such an affordable yet modern wireless communication system can enable telemedicine service to bring worldwide clinical expertise to the point of care in low-resource regions. The authors in~\cite{Martinez2002} discussed the role of ICT in the context of their \textit{EHAS} (Hispano-American Health Link) program. The burden on healthcare professionals arising due to the lack of ICT tools was highlighted in this work too. The authors discussed the deployment of custom email servers, working over the Very High Frequency (VHF) band for communication, to help healthcare establishments in low-resource regions of Peru. The recent roll-outs of advanced telecommunication technologies such as 5G, if available in low-resource regions, can significantly enhance healthcare service delivery in those regions~\cite{magaia2021artificial,shaylor2021tale}.

Some specific features of ICT software advancements e.g. features in the database have also been advantageously applied in some works. In~\cite{Puustjaervi2011}, the authors discussed their software tool \textit{Health Agent} which is useful to automate telemedicine in rural areas. The functionalities of a relational database system, e.g. SQL triggers, are utilized for case-based alarming in the proposed solution. The data model for proper storage and access to health information is also discussed. These data model development utilizes relevant advances from the larger internet technology developments.

\subsection{ICT for healthcare knowledge and language tools}

The role of ICT tools in knowledge management and training of health professionals has also been a subject of discussion in some previous works. Proper knowledge management and affordable training are important for healthcare in low-resource regions because of the general lack of trained professionals and knowledge resources. In~\cite{Emmerling2014}, the authors proposed a digital library solution to assist healthcare in low-resource regions. The library consists of reading materials on healthcare physiology, tools maintenance, operations, etc. This information is made available in the local language using an automatic language translation service. The authors in~\cite{Ruijter2008} presented an open learning platform called \textit{Openlearn} to provide free training materials on healthcare management. A free training platform could in particular be beneficial for personnel in low-resource healthcare settings who cannot afford or access training with high costs. In~\cite{Bondale2013}, the authors presented different ICT innovations to support primary healthcare in India (relevant for low-resource regions also). Replacing manual healthcare information management with automated (ICT) solutions for efficiency is highlighted. Similarly, software-as-a-service deployments and mobile applications for improved connectivity of mobile midwives are presented to make the maternity healthcare system more efficient. For any innovation, the availability of multi-language tools, e.g., owing to open-source tools and technologies for language translation~\cite{opennmt}, is considered important. In general, the need for healthcare information and user interface tools in local languages is identified as very crucial for healthcare services in low-resource regions where English might not be the primary language of the majority. These intuitive user interfaces and language translation tools are directly linked to ICT innovations. In~\cite{Amararachchi2013}, the authors discussed the role of an ICT-based knowledge management system for improved healthcare access in low-resource regions of Srilanka. The need for a knowledge management system for making healthcare pervasive in low-resource regions is made clear. Still, some problems like the skeptical attitude of some clinical professionals in using ICT resources, general unavailability of good systems and software, etc. were raised. New propositions should consider user-centric ICT propositions that are adapted to local needs to improve the trust and usability of developed propositions.   

\subsection{Section summary}

To summarize, several propositions have used ICT advancements for improving healthcare access and affordability in low-resource regions. The open-source software tools, increasing access to network/communication infrastructure, easy access to health information library, etc., are all ICT advancements also helping healthcare in low-resource regions.

Some studies have shown the direct relation between ICT advancements and feasible healthcare innovations. In~\cite{ali2019}, the authors studied the quantitative relationship between IT infrastructure and telemedicine readiness in the context of healthcare systems in Libya. A positive correlation of 0.44 was obtained, highlighting the direct impact of ICT infrastructure on enabling new healthcare innovations in low-resource regions. Besides the general availability of ICT infrastructures, the authors also highlighted the need for associated trained professionals for its operation and maintenance. In~\cite{Montalban2008}, the authors interviewed several healthcare professionals working in low-resource regions of the Philippines. A need for proper access to clinical practice guidelines, health news, and events, health education policies, etc. were identified from the interviews. This information is made accessible with ICT infrastructures, and thus the role of ICT for healthcare in low-resource regions is highlighted.

\section{Mobile Health}
\label{section:mobile_health}

Mobile Health (mHealth) refers to the practice of health solutions and services supported by mobile devices such as mobile phones, tablets,  PDAs, and wearable devices. The mHealth field has emerged as a sub-segment of eHealth. The services generally considered under mHealth include the use of mobile devices in collecting community and clinical health data, delivery of healthcare information to practitioners, researchers, and patients, real-time monitoring of the patient, and direct provision of care (via mobile telemedicine).  

mHealth has emerged in recent years to be suitable for developing countries, resulting from the rapid rise of mobile phone penetration in the low-resource regions also. It can enable wider healthcare access to a larger segment of the population in low-resource regions by improving the capacity of health systems. Here we discuss some works that have focused on mHealth. We also list several health-related mobile applications that could be relevant for deployment in low-resource regions or regions with limited access to conventional healthcare services.   

The authors in~\cite{Sinharay2016} highlight how mobile devices can be used as medical devices. The use of the available capacity of smartphones, e.g., the onboard sensors, web connectivity, and powerful processing unit, has made these mobile devices not just a medium of communication but also a medical device. Providing proof of concept with a digital stethoscope, the importance of available infrastructure to cope with critical diseases related to heart patients has been discussed. In the low-resource settings, mHealth could be the only way to offer services that would otherwise be unavailable~\cite{Brown2014}. Mobile phones with affordable network connectivity could be a vector for enhancing the continuity of care before the diagnosis and after the hospitalization for a continuous follow-up. mHealth applications allowing networking and video conferencing could help to increase efficiency (better logistics, more contact, etc.), quality of patient care (e.g., due to more follow-up even if remote), and patient satisfaction. There might be some constraints on mHealth applications when considering their usage in low-resource settings. The authors in~\cite{Hebert2016}, for example, investigated the effectiveness of mobile apps in rural areas with low literacy. Visual-based apps were recommended for better meeting the needs in low-resource regions.

We list some example mobile/smartphone applications (mobile apps) in Table~\ref{table:App_diagnosis},\ref{table:App_record} and \ref{table:App_life} that could be relevant for healthcare in low-resource settings. We categorized the applications based on the target complications they address. Furthermore, we classified the apps based on their feature type. These apps can be used either directly by the patients and families, or by midwives and doctors.

\begin{table}

\centering 

\caption{Some example smartphone applications developed for health monitoring and diagnosis. These applications are low-cost and could be useful to aid healthcare services in low-resource regions.}

\label{table:App_diagnosis}
\begin{tabular}{|p{5cm}|p{9cm}|}
\hline 
\textbf{Name} & \textbf{Target Complication} \\ 
\hline 
iASHA \cite{Mukherjee2012}&	Pregnancy (A decision support system to enable health workers to provide maternity healthcare services efficiently and transparently.\\
\hline
mBody Health \cite{Hebert2016} &	Diagnose disabilities \\
\hline
Mobile Stethoscope \cite{Sinharay2016} &	Monitor Heart rate to diagnose heart disease\\
\hline
Cardiax Mobile ECG~\cite{Cardiax} & 12 channel personal ECG system to diagnose and monitor heart condition \\
\hline
Runtastic Heart rate~\cite{runtastic} &	Measures heart rate on a real-time basis \\
\hline
Heart rate monitor~\cite{HeartRate} &	Checks heart rate on a real-time basis\\
\hline
Blood pressure watch~\cite{BloodPressureWatch} &	Collects, tracks, analyzes, and shares blood pressure data \\
\hline
Finger blood pressure~\cite{FingerBloodPressure}	& Measures blood pressure based on imaging of fingertip \\
\hline
Finger Print Thermometer~\cite{FingerPrintThermometer} &	Determines body temperature based on contact at fingerpoint\\
\hline
Body temperature~\cite{BodyTemperature} &	Keep track of body temperature and identify severity\\
\hline
iOximeter~\cite{ioximeter} &	Pulse rate calculator \\
\hline
Eye care plus~\cite{eyecareplus} &	Test and monitor vision\\
\hline
Test your hearing~\cite{testyourhearing} &	Test hearing \\
\hline
uHear~\cite{uhear} &	Self-assessment of hearing\\
\hline

\end{tabular} 
\end{table}

\begin{table}

\centering 

\caption{Some example smartphone applications for healthcare record-keeping. Measurement and tracking of health through record keeping are crucial aspects of healthcare delivery.}

\label{table:App_record}
\begin{tabular}{|p{5cm}|p{9cm}|}
\hline 
\textbf{Name} & \textbf{Target Complication} \\ 
\hline 
iASHA \cite{Mukherjee2012} &	Pregnancy\\
\hline
mBody Health \cite{Hebert2016} & Monitor disabilities\\
\hline
Mobile Stethoscope \cite{Sinharay2016} &	Monitor heart disease\\
\hline
Health assistant~\cite{healtassistant} &	Monitor BP, weight, water, temp, glucose, etc\\
\hline
Asthma tracker and log~\cite{asthmatracker} & Asthma tracker\\
\hline
Diabetes Diagnostics~\cite{diabetesdiagnostics} & Diagnose diabetes subtypes\\
\hline
\end{tabular} 
\end{table}

\begin{table}
\centering

\caption{Example smartphone applications related to lifestyle and healthy behaviors. The concept of healthcare is a continuum that includes the day-to-day life of an individual and thus applications helping in lifestyle and healthy behaviors are relevant for improving healthcare overall, including in the low-resource regions.}
\label{table:App_life}
\begin{tabular}{|p{5cm}|p{9cm}|}
\hline 
\textbf{Name} & \textbf{Target Complication} \\ 
\hline 
Health assistant~\cite{healtassistant} &	BP, weight, water, temp, glucose, etc. \\
\hline
Water Your body~\cite{wateryourbody} &	Reminds user to drink water \\
\hline
Medisafe Meds \& Pill Reminder~\cite{medisafe} & Reminds user of medication times\\
\hline
Dosecast indication Reminder~\cite{dosecast} & Drug management\\
\hline
MyFitnessPal~\cite{myfitnesspal} & Keep track of food habits\\
\hline
SleepCycle~\cite{sleepcycle} & Keep record of sleep time\\
\hline
FitBit~\cite{fitbit}	 & Track daily goal and progress over time for steps, distance, calories burned and more.\\
\hline
\end{tabular} 
\end{table}

\section{Emerging Technologies}
\label{section:emergin_technologies}

Some of the emerging technologies like artificial intelligence (AI), Internet of Things (IoT), Blockchain, etc., have already shown great promise for healthcare. The technologies also have a role in making healthcare inclusive and affordable. In this section, we review some works where emerging technologies for healthcare have been discussed. Though not always explicitly stated, these works have the potential translation of their usage for increasing the outreach of healthcare in low-resource settings.

\subsection{AI for Healthcare}

AI has shown its potential in a broad set of use cases. Much of this success is due to the availability of data, easily available computing power, and an improved understanding of different AI algorithms. Machine learning, which is concerned with learning from data, is one of the enablers of AI. Many advancements in healthcare due to AI have been brought by machine learning advances. In recent years, deep-learning, a type of machine learning based on multi-layered neural network architectures, has provided even clinician-grade solutions, e.g., to analyze medical images and physiological signals.  For the healthcare application, AI has a role in several areas like automated diagnosis, clinical decision support, clinical workflow optimization, resource utility maximization, etc.

\begin{figure}
\label{figure:generic_ai_flow}
\includegraphics[width=0.9\columnwidth]{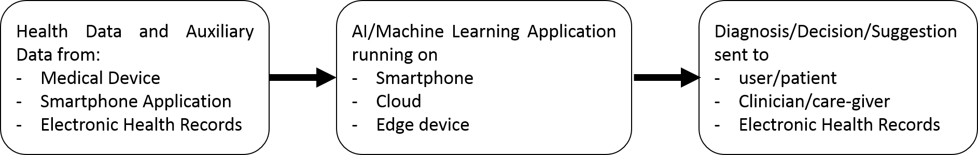}
\caption{Generic data flow for AI-enabled healthcare diagnosis and decision support applications. The data collected from a sensor front-end, or retrospective data from electronic health records, are sent to a computing platform running an AI application. This computing platform could be a smartphone, cloud, edge device, or any other computing device. AI application in the computing platform takes the input data and provides diagnosis and decision support as output. The output is provided either to the user (e.g., for smartphone applications) or to the clinician and caregiver. The output can also be linked to the patient’s/user’s Electronic Health Records to keep track of his/her health.}
\end{figure}

The use of AI for diagnosis has been one of the most explored application areas. The typical data flow for such a solution is shown in Figure \ref{figure:generic_ai_flow}. Data from a (multi-) sensor front-end is consumed by an AI application to produce diagnosis and decision support outputs. Based on a usage area, AI for diagnosis can be categorized into: (i). Inside-hospital application and (ii). Outside-hospital application. 

\subsubsection{Inside-hospital applications}

The inside-hospital applications use AI for aiding diagnosis for day-to-day clinical workflows. Several innovations in this area are, for example, in the field of radiology. An example application is tumor/nodule detection in the lung CT scan~\cite{rubin2015characterizing}. In \cite{Hosny2018}, the authors identify several application areas where AI could assist in radiology-based diagnosis. Some of those areas are thoracic imaging for lung nodule detection, abdominal and pelvic imaging for liver lesions detection, colonoscopy for colonic polyps detection, brain imaging for brain tumors detection, etc. While the radiology equipment might be a significant cost burden for low-resource settings, at least the automation of (pre-)diagnosis could offset the cost required for well-trained radiologists. Besides radiology, other application areas of AI in clinical diagnosis and decision making include for example hospital discharge or readmission prediction
\cite{Avati2018,ahn2021machine,chen2022machine} using available clinical data, or arrhythmia detection using ECG (Electrocardiograph) signal \cite{Hannun2019,jo2021detection, ccinar2021classification}.

\subsubsection{Outside-hospital applications}

The outside-hospital application categories are concerned with taking the (early/pre-) diagnosis to the ambulatory settings of free-living. A representative work is presented by the authors in~\cite{Esteva2017}. The authors propose a camera-based system to provide a clinical-grade diagnosis of skin cancer. A motivation for such a system is the possibility to extend the reach of dermatologists outside of the traditional setting and enable low-cost healthcare on top of ubiquitously available smartphones.  The authors of~\cite{Voisin2018} describe a deep-learning-based atrial fibrillation detection solution using photoplethysmography (PPG) signals in an ambulatory setting. These example applications show how the power of AI can help take the clinical diagnosis to ambulatory settings. Such a step is crucial to bringing quality diagnosis solutions to low-resource areas. We refer to the review works in~\cite{jiang2017artificial, yu2018artificial, reddy2019artificial, vaananen2021ai, shaheen2021applications}  for the application of AI/machine learning in healthcare areas.

Besides diagnosis and clinical decisions, AI could be advantageous in other aspects of the healthcare delivery chain. Any AI-based solution to improve the healthcare delivery chain can translate to a direct benefit towards healthcare for low-resource settings. The benefits can be cost reduction, improved efficiency, the requirement of lesser resources and manpower, etc. An example discussing the use of AI to improve the entire healthcare delivery chain is the work of~\cite{Doshi2017}. The authors discuss how AI can be used for the diagnosis and treatment of Tuberculosis (TB) by improving all the aspects of the healthcare chain. AI technologies can help assure medication adherence using automated gesture detection technologies (e.g., from video surveillance). AI can be useful for the whole program management of TB containment and treatment in an area e.g., in the context of resource planning and monitoring. For an application like TB, raising awareness of the patients and population is also very important. For this purpose, using AI for the automated generation of educational audio-visual materials can be helpful. Other works have also outlined how AI and optimization can be helpful in aspects other than core clinical tasks like diagnosis and treatment. The authors in~\cite{Brunskill2010} have proposed optimized planning for scheduling the work of Community Health workers where AI can play a role in the optimized scheduling. AI can be used in different health resource planning works like referrals, management of immunization programs, and several population/public health initiatives. Such efficient planning can be crucial when it comes to reducing the cost of healthcare in low-resource settings and improving throughput/efficiency. The authors in~\cite{mesko2018will} also discuss how AI can have a role in healthcare in low-resource settings for various functions like diagnosis, administration, resource planning, monitoring, etc. Specific example deployments like AI-based chat-bots to support medication has been presented. The challenge concerning the cost of AI-based solutions is raised but it has also been noted that overall cost reduction can be expected, justifying the fit of AI-based systems for healthcare in low-resource settings. An advantage for AI deployment in low-resource settings could be the relative policy-level ease of deployment. The example of clinical drone usage in Rwanda has been presented to show this point.

To summarize, the growth and increasing adoption of AI can prove to be very helpful for increasing the outreach of healthcare in low-resource settings. AI as a technology is being widely investigated to improve clinical diagnosis and decision making, either inside the hospital or outside. Besides diagnosis, AI can also assist in reducing costs and improving efficiency for other aspects of the healthcare chain. Though the initial cost factor and required technological knowledge for solution deployment can seem like a hindrance to getting AI technologies to the low-resource settings, AI provides several other benefits which could pay off overall.

\subsection{Internet of Things (IoT) for healthcare}

Connectivity and computational power are getting cheaper and more ubiquitous. This has led to a myriad of devices around us being more intelligent and inter-connected. The connectedness of everyday items and devices around us, which also get smart and intelligent, is referred to as the Internet of Things (IoT). Think of an automatic watering unit for the plant enamored by environmental sensors that we can monitor and control using a mobile application. Or a refrigerator that has an inbuilt camera to detect missing groceries and update the shopping list on our smartphone. IoT is already a reality today, and only growing in terms of scale and usage. 

Healthcare is one of the areas where IoT can enable new solutions. Most of the research work on IoT for healthcare has mHealth or a wireless body area network (WBAN) application deployment scenarios. Wearable physiological sensors connected to the cloud and made available for remote access by the clinician is a typical application scenario. This, for example, has been discussed by~\cite{Krishnan2016, Ganesh2016,Raj2017,Divakaran2017,Prakashan2017, Zilani2018,ryu2021comprehensive,sriraam2021wearable,dese2022low}. Innovative deployments have been discussed in works like that of~\cite{Sarkar2017} where the authors present a health-kiosk available for healthcare in remote regions of India. These kiosks consist of multiple physiological sensors which can make the data available for remote access through cloud services. The authors in~\cite{Pathinarupothi2018} discuss the possibility of an intelligent edge device within an IoT deployment of personal physiological sensors in low-resource settings. The edge device can provide various useful functionalities like smart data filtering to only forward relevant information to the clinicians in a remote location. It can also automatically detect any health criticality among the monitored population. Such deployments increase the (virtual) outreach of clinicians. An interesting extension of this smart edge-based IoT deployment is the possibility to monitor the entire (sub) population in the region, leading to population health analytics solutions. Besides personal physiological sensors monitoring the health of individuals, other devices can be a part of the IoT like a visitor counter device at the local health center, an EMR server of the health center, etc.  

A review of IoT for healthcare, though not specific to low-resource settings, has been presented in the work of~\cite{Islam2015}. The authors discuss various IoT use-cases for healthcare applications. The potential of large-scale IoT for applications in community health monitoring discusses concepts like interactions between multiple body area networks. An interesting use case is IoT's use for emergency healthcare services during a disaster. In low-resource settings where the reach of standard healthcare services is limited,  connectivity of devices could help monitor and plan resource utilization, casualties mapping, etc. One other use case discussed is the possibility to track population-level information on adverse drug reactions. This can immediately be extended to applications specific to low-resource settings e.g., tracking outbreaks and the spread of specific diseases related to low-quality water supply, for example. IoT can also be applied for awareness campaigns by providing a platform for easy dissemination of information.

\subsection{Blockchain technology for healthcare}

One of the recent technological advancements with huge potential in the healthcare sector, positively impacting the increase of the outreach in low-resource settings, is blockchain technology. Blockchain technology enables a way to keep a record of the transaction without a need for an intermediary. This mechanism opens up various applications in healthcare that would otherwise have been difficult to deploy. As blockchain provides a trust mechanism between multiple parties, one of the most explored applications of blockchain in healthcare is the sharing of data. This application is very relevant for low-resource settings for two reasons. First, as healthcare services and providers are not resourceful, sharing data between different entities is difficult. E.g., healthcare facilities might not have enough software and resources to open up their data to third parties without incurring a major security and privacy threat. Second, the new applications that can be enabled with easy data sharing would be crucial in low-resource settings to enable applications like automated diagnosis, telemedicine, or any general mHealth applications.  

The authors in~\cite{weiss2017blockchain} discuss how blockchain could provide privacy, traceability, value-management, and trust mediation for enabling mHealth applications in South Africa where the requirements for healthcare access extends to some deep rural settings also. In~\cite{kamau2018blockchain}, the authors propose the use of blockchain to manage distributed information in EMR, providing the security and trust mechanism for sharing information. This is discussed in the context of managing EMR for cancer patients. Given the lack of inter-connectivity or secure direct connectivity between different data/information nodes, blockchain can provide a secure gluing layer. The benefit of decentralization to improve healthcare in low-resource countries has been discussed in~\cite{ayubblockchain}. Though not explicitly mentioned, blockchain can be the technology upon which a distributed healthcare infrastructure is built. The use of blockchain for data sharing providing functionalities like patient pseudo-anonymity, workflow automation, data integrity, accountability, etc. has been presented in the paper of~\cite{theodouli2018design}. Similarly, the authors of~\cite{liang2017integrating} present how the blockchain technology in a mHealth application provides the solution for the right data ownership, security of the shared data, convenience of sharing, etc. The authors in~\cite{alexaki2018blockchain} uses blockchain and smart contracts as a mediator to enable data-sharing across the providers (ensuring interoperability) to make healthcare more cost-effective and sustainable. In the MedicalChain whitepaper~\cite{medicalchain}, the use of blockchain and smart contract platform enabled by blockchain for data exchange for solutions like telemedicine is discussed. The use of blockchain in healthcare, in general, has been reviewed in several works~\cite{lam018applications,holbl2018systematic,mcghin2019blockchain,attaran2022blockchain,sookhak2021blockchain}.

As more works are pursued on the usage of blockchain in various aspects of healthcare like diagnosis, treatment, clinical workflow, logistics, prescription management, logistics management, etc., it would become clearer what promises this new technology will bear for healthcare in low-resource settings. From what we have already started to see, blockchain technology has the potential to bring positive changes to healthcare in a low-resource setting also.

\section{Discussion}
\label{section:discussion}

Quality healthcare access should be available for all. However, due to higher costs associated with healthcare resources and infrastructures, we are yet to reach the utopian dream of getting everyone on the radar of good healthcare access. In this work, we reviewed various works that have been geared towards making healthcare widely accessible in low-resource settings.

A major effort to control healthcare costs concerns reducing the price of costly healthcare devices. This is becoming feasible due to the reducing cost of semiconductors, the pricing reduction possible with improved mass manufacturing, and the investigation of new designs with the cost factor taken into consideration. We found that several efforts have been focused on building an add-on for the smartphone, so that the smartphone itself can act as a healthcare device, for example for early diagnosis. These works rely on the wider availability of smartphones in a low-resource setting. The reach of smartphones has been growing very rapidly even in the rural and low-resource settings~\cite{Karlsson2018} and thus the smartphone as an add-on could be transformative. The end-to-end healthcare delivery chain requires several devices. Therefore, reducing the costs of these devices is a major step, and a crucial one, towards universal healthcare access. However, the pressure of cost reduction should not compromise the device quality. The quality reduction might lead to an adverse outcome for the patient. The concern of quality is especially relevant for low-resource settings because these areas generally lack good regulations for devices and/or are lacking in trained personnel to enforce quality guidelines or maintenance~\cite{panta2020compliance, thapa2022effect}. Another point to be noted is the overall representation of the device cost in the financial equation of the healthcare delivery. It might as well be that the device costs represent only a small portion of the overall healthcare costs, in a given care setting. Also, the device costs might be only a one-time capital expenditure which is overshadowed by the recurring operational costs. Therefore, all efforts toward affordability of the healthcare device have to take the following factors into account in further investigations: \textit{What fraction of the healthcare delivery cost is represented by the device costs?}, \textit{Can the cost reduction be done without reducing quality?}, \textit{How is the distribution of the capital expenditure and operation cost of the device? And how does new low-cost design impact these two factors?}.

A development for increased healthcare access in a low-resource setting is also the parallel development, or rather an explosion, of IT tools and technologies. Software-based services and telecommunication infrastructure can provide a carrier for innovations in healthcare targeted at low-resource settings. If we think of novel healthcare offerings like telehealth, these offerings have a good landing spot in the context of healthcare for low-resource settings. Several studies have already investigated how the availability of a new generation of telecommunication technology is going to help revolutionize healthcare~\cite{singh2021potential, kumar2022importance, minervini2022teledentistry, el2022telehealth}. A bit of that revolution will be shared by healthcare for low-resource settings. Another development is in the field of open-source IT tools. An example is OpenEMR which has been positioned to provide a reliable EMR solution. Open-source solutions such as OpenEMR could fill the void created when the healthcare organizations in the low-resource settings cannot afford proprietary EMR solutions. Needless to say, when it comes to the use of open-source software and tools for healthcare, utmost care should still be provided to various aspects like quality, privacy, etc. It would be unfortunate if the cost factor of a software tool is used to overlook any impending harms like compromised patient privacy or even patient safety. The efforts of healthcare for low-resource settings should embrace developments in the open-source world but with caution due to the criticality of the healthcare sector in general. Even though some recent studies have found the privacy and security concern to be of less significance to eHealth users in low-resource settings~\cite{archer2021ehealth}, this might only reflect the notion that having the resources/infrastructure is of the highest significance in regions where these are missing. Some studies have, however, found that patients in low-resource regions are concerned by the unauthorized secondary use of their protected health information~\cite{adu2019individuals}. Greater awareness about the implications of security and privacy breaches on one's health information will likely increase the concern of patients on these factors.

A major carrier for increased healthcare access in low-resource could be smartphones. The smartphone revolution will likely usher changes in healthcare delivery, with a positive impact on cost reduction and affordability for people in low-resource settings. In section~\ref{section:mobile_health}, we reviewed several healthcare-related smartphone applications. Various aspects like diagnosis, record-keeping, improving lifestyle towards a healthier life, etc., are covered by existing and upcoming applications. The smartphone has inbuilt sensors like cameras and microphones that could work as health sensors, something which has been receiving a lot of attention lately. For example, cameras could extract heart rate variability and other physical health parameters~\cite{hasan2022camsense}. Similarly, the microphone can be used in speech-related applications, for example, to assess sociability for depression severity estimation~\cite{lamichhane2022econet}. The smartphone-based applications that we reviewed in Section~\ref{section:mobile_health} covered a wide range of clinical conditions for cardiac care, optical care, pregnancy monitoring, thermal monitoring, and many more. The smartphone revolution can thus have a wide impact on healthcare. Though the concept of a simple application to diagnose a disease at home sounds lucrative, it is important to ensure the accuracy of such applications by validating with current standard devices. Similarly, most of the applications will be customer-facing. Therefore it is also important to think about how the accessibility is not impeded by factors like the literacy level of people in a particular region. 
Just like the mature technologies such as smartphones and telecommunications which have been helping improve healthcare access in low-resource settings, one can also remain upbeat with many upcoming technologies. In Section~\ref{section:emergin_technologies}, we identified some technologies like IoT, Blockchain, and Artificial Intelligence which will all bring in another wave of revolution in healthcare. We reviewed different literature where the authors are investigating how new technologies will change healthcare services. It might not be too early to speculate that these new technologies will further help improve healthcare access in low-resource settings. Some recent works concur, for example with the potential of blockchain technology to reduce health disparities and improve global health~\cite{thomason2021blockchain}.

\bibliographystyle{unsrt}  
\bibliography{healthcare_paper} 

\begin{thebibliography}{100}

\bibitem{WHO2018}
Ageing and health, world health organization (who).
\newblock
  \url{https://www.who.int/news-room/fact-sheets/detail/ageing-and-health},
  2018.
\newblock Last accessed: 2022-05-14.

\bibitem{Rowe2016}
John~W. Rowe, Terry Fulmer, and Linda Fried.
\newblock Preparing for better health and health care for an aging population.
\newblock {\em JAMA}, 316:1643--1644, Oct 2016.

\bibitem{Gupta2015}
Shailvi Gupta, MDa, Reinou Groen, MDb, Patrick Kyamanywac, Emmanuel Ameh, MDd,
  Mohamed Labib, MDe, Damian Clarke, MDf, Peter Donkor, MDg, Miliard Derbew,
  MDh, Rachid Sani, MDi, Thaim Kamara, MDj, and Adam Kushner.
\newblock Surgical care needs of low-resource populations: an estimate of the
  prevalence of surgically treatable conditions and avoidable deaths in 48
  countries.
\newblock {\em The Lancet}, 385, April 2015.

\bibitem{UN2020}
Transforming our world : the 2030 agenda for sustainable development
  a/res/70/1.
\newblock \url{https://www.refworld.org/docid/57b6e3e44.html}, 2015.
\newblock Last accessed: 2022-05-14.

\bibitem{WHO2012}
WHO.
\newblock Prevention and control of ncds: Guidelines for primary health care in
  low-resource settings, 2012.

\bibitem{WHO2018_1}
The 2022 update, global health workforce statistics, world health organization,
  geneva.
\newblock
  \url{https://www.who.int/data/gho/data/themes/topics/health-workforce}, 2022.
\newblock Last accessed: 2022-05-14.

\bibitem{AlShamsi2017}
Mustafa Al-Shamsi.
\newblock Addressing the physicians' shortage in developing countries by
  accelerating and reforming the medical education: Is it possible?
\newblock {\em Journal of advances in medical education \& professionalism},
  5:210--219, Oct 2017.

\bibitem{Mullan2007}
Fitzhugh Mullan and Seble Frehywot.
\newblock Non-physician clinicians in 47 sub-saharan african countries.
\newblock {\em Lancet (London, England)}, 370:2158--63, Dec 2007.

\bibitem{Siddiqi2005}
Kamran Siddiqi and James~N. Newell.
\newblock Putting evidence into practice in low-resource settings.
\newblock In {\em Bulletin of the World Health Organization}, volume~83, page
  882, Switzerland, Dec 2005.

\bibitem{Heigener2014}
D.~Heigener, C.~Knapp, R.~Viola, Uzochukwu~Uzoma Aniebue, and Tonia~Chinyelu
  Onyeka.
\newblock Ethical, socioeconomic, and cultural considerations in gynecologic
  cancer care in developing countries.
\newblock {\em International Journal of Palliative Care}, 2014:141627, 2014.

\bibitem{noordam2015use}
Aaltje~Camielle Noordam, Yolanda Barber{\'a}~La{\'\i}nez, Salim Sadruddin,
  Pabla~Maria van Heck, Alex~Opio Chono, Geoffrey~Larry Acaye, Victor Lara,
  Agnes Nanyonjo, Charles Ocan, and Karin K{\"a}llander.
\newblock The use of counting beads to improve the classification of fast
  breathing in low-resource settings: a multi-country review.
\newblock {\em Health Policy and Planning}, 30(6):696--704, 2015.

\bibitem{Karlsson2018}
M~Karlsson.
\newblock Accelerating affordable smartphone ownership in emerging markets.
  walbrook: Gsma., 2018.

\bibitem{wahl2018}
Brian Wahl, Aline Cossy-Gantner, Stefan Germann, and Nina~R Schwalbe.
\newblock Artificial intelligence (ai) and global health: how can ai contribute
  to health in resource-poor settings?
\newblock {\em BMJ global health}, 3(4):e000798, 2018.

\bibitem{dash2020}
Satya~Prakash Dash.
\newblock The impact of iot in healthcare: Global technological change \& the
  roadmap to a networked architecture in india.
\newblock {\em Journal of the Indian Institute of Science}, pages 1--13, 2020.

\bibitem{upadhyaya2018}
Pranita Upadhyaya, Sanjib~Kumar Upadhyay, Bishikha Subedi, Bhawana Subedi, and
  Asmita Gaire.
\newblock Revolutionizing healthcare systems of a developing country using
  blockchain.
\newblock In {\em 2018 IEEE International Conference on Computational
  Intelligence and Computing Research (ICCIC)}, pages 1--6. IEEE, 2018.

\bibitem{compton2018}
Bruce Compton, David~M Barash, Jennifer Farrington, Cynthia Hall, Dale Herzog,
  Vikas Meka, Ellen Rafferty, Katherine Taylor, and Asha Varghese.
\newblock Access to medical devices in low-income countries: addressing
  sustainability challenges in medical device donations.
\newblock {\em NAM Perspectives}, 2018.

\bibitem{Byrne2017}
David~M. Byrne, Stephen~D. Oliner, and Daniel~E. Sichel.
\newblock {How Fast are Semiconductor Prices Falling?}
\newblock Finance and Economics Discussion Series 2017-005, Board of Governors
  of the Federal Reserve System (U.S.), January 2017.

\bibitem{Guthrie2012}
B.~J. Guthrie.
\newblock Low cost blood glucose meters as an appropriate healthcare technology
  for developing countries.
\newblock In {\em 7th International Conference on Appropriate Healthcare
  Technologies for Developing Countries}, pages 1--3, 2012.

\bibitem{mohan2020management}
Viswanathan Mohan, Kamlesh Khunti, Siew~P Chan, F~Fadlo~Filho, Nam~Q Tran,
  Kaushik Ramaiya, Shashank Joshi, Ambrish Mithal, Ma{\"\i}mouna~N Mbaye,
  Nemencio~A Nicodemus, et~al.
\newblock Management of type 2 diabetes in developing countries: balancing
  optimal glycaemic control and outcomes with affordability and accessibility
  to treatment.
\newblock {\em Diabetes Therapy}, 11(1):15--35, 2020.

\bibitem{cho2018idf}
NH1 Cho, JE~Shaw, Suvi Karuranga, Y~da Huang, JD~da~Rocha~Fernandes,
  AW~Ohlrogge, and B~Malanda.
\newblock Idf diabetes atlas: Global estimates of diabetes prevalence for 2017
  and projections for 2045.
\newblock {\em Diabetes research and clinical practice}, 138:271--281, 2018.

\bibitem{nikpour2021innovative}
Nazaneen Nikpour~Hernandez, Samiha Ismail, Hen Heang, Maurits van Pelt, Miles~D
  Witham, and Justine~I Davies.
\newblock An innovative model for management of cardiovascular disease risk
  factors in the low resource setting of cambodia.
\newblock {\em Health Policy and Planning}, 36(4):397--406, 2021.

\bibitem{Pimentel2014}
M.~A.~F. Pimentel, M.~D. Santos, M.~A. Maraci, C.~Arteta, J.~S. Domingos, D.~A.
  Clifton, and G.~D. Clifford.
\newblock A \$5 smart blood pressure system.
\newblock In {\em Appropriate Healthcare Technologies for Low Resource Settings
  (AHT 2014)}, pages 1--4, 2014.

\bibitem{Sinharay2016}
A.~Sinharay, D.~Ghosh, P.~Deshpande, S.~Alam, R.~Banerjee, and A.~Pal.
\newblock Smartphone based digital stethoscope for connected health -- a direct
  acoustic coupling technique.
\newblock In {\em 2016 IEEE First International Conference on Connected Health:
  Applications, Systems and Engineering Technologies (CHASE)}, pages 193--198,
  2016.

\bibitem{jain2021development}
Agam Jain, Roshan Sahu, Arohi Jain, Thomas Gaumnitz, Prayas Sethi, and Rakesh
  Lodha.
\newblock Development and validation of a low-cost electronic stethoscope: Diy
  digital stethoscope.
\newblock {\em BMJ Innovations}, 7(4), 2021.

\bibitem{chorba2021deep}
John~S Chorba, Avi~M Shapiro, Le~Le, John Maidens, John Prince, Steve Pham,
  Mia~M Kanzawa, Daniel~N Barbosa, Caroline Currie, Catherine Brooks, et~al.
\newblock Deep learning algorithm for automated cardiac murmur detection via a
  digital stethoscope platform.
\newblock {\em Journal of the American Heart Association}, 10(9):e019905, 2021.

\bibitem{ali2021protocol}
Fatima Ali, Babar Hasan, Huzaifa Ahmad, Zahra Hoodbhoy, Zainab Bhuriwala,
  Muhammad Hanif, Shahab~U Ansari, and Devyani Chowdhury.
\newblock Protocol: Detection of subclinical rheumatic heart disease in
  children using a deep learning algorithm on digital stethoscope: a study
  protocol.
\newblock {\em BMJ Open}, 11(8), 2021.

\bibitem{grooby2022prediction}
Ethan Grooby, Chiranjibi Sitaula, Kenneth Tan, Lindsay Zhou, Arrabella King,
  Ashwin Ramanathan, Atul Malhotra, Guy~A Dumont, and Faezeh Marzbanrad.
\newblock Prediction of neonatal respiratory distress in term babies at birth
  from digital stethoscope recorded chest sounds.
\newblock {\em arXiv preprint arXiv:2201.10105}, 2022.

\bibitem{chen2021toward}
Yuhan Chen, Michael~D Wilkins, Jeffrey Barahona, Alan~J Rosenbaum, Michael
  Daniele, and Edgar Lobaton.
\newblock Toward automated analysis of fetal phonocardiograms: Comparing
  heartbeat detection from fetal doppler and digital stethoscope signals.
\newblock In {\em 2021 43rd Annual International Conference of the IEEE
  Engineering in Medicine \& Biology Society (EMBC)}, pages 975--979. IEEE,
  2021.

\bibitem{Lawn2006}
J.~Lawn, J.~Wyatt, D.~Woods, H.~Bezuidenhout, J.~Lawn, J.~Wyatt, D.~Woods, and
  H.~Bezuidenhout.
\newblock Does fetal heart rate count? developing a low cost, alternative
  powered doppler fetal heart monitor for use in low resource high mortality
  settings.
\newblock In {\em 2006 The 4th Institution of Engineering and Technology
  Seminar on Appropriate Healthcare Technologies for Developing Countries},
  pages 155--161, 2006.

\bibitem{Cardillo2014}
H.~Cardillo, J.~Kohler, E.~Kriner, and K.~Mehta.
\newblock Applications of wood's lamp technology to detect skin infections in
  resource-constrained settings.
\newblock In {\em IEEE Global Humanitarian Technology Conference (GHTC 2014)},
  pages 548--554, 2014.

\bibitem{Haney2017}
K.~Haney, P.~Tandon, R.~Divi, M.~R. Ossandon, H.~Baker, and P.~C. Pearlman.
\newblock The role of affordable, point-of-care technologies for cancer care in
  low- and middle-income countries: A review and commentary.
\newblock {\em IEEE Journal of Translational Engineering in Health and
  Medicine}, 5:1--14, 2017.

\bibitem{Gouwanda2016}
D.~Gouwanda, A.~A. Gopalai, and B.~H. Khoo.
\newblock A low cost alternative to monitor human gait temporal
  parameters-wearable wireless gyroscope.
\newblock {\em IEEE Sensors Journal}, 16(24):9029--9035, 2016.

\bibitem{powell2021investigating}
Dylan Powell, Mina Nouredanesh, Samuel Stuart, and Alan Godfrey.
\newblock Investigating the ax6 inertial-based wearable for instrumented
  physical capability assessment of young adults in a low-resource setting.
\newblock {\em Smart Health}, 22:100220, 2021.

\bibitem{Palmius2014}
N.~Palmius, M.~Osipov, A.~C. Bilderbeck, G.~M. Goodwin, K.~Saunders, A.~Tsanas,
  and G.~D. Clifford.
\newblock A multi-sensor monitoring system for objective mental health
  management in resource constrained environments.
\newblock In {\em Appropriate Healthcare Technologies for Low Resource Settings
  (AHT 2014)}, pages 1--4, 2014.

\bibitem{lamichhane2020patient}
Bishal Lamichhane, Dror Ben-Zeev, Andrew Campbell, Tanzeem Choudhury, Marta
  Hauser, John Kane, Mikio Obuchi, Emily Scherer, Megan Walsh, Rui Wang, et~al.
\newblock Patient-independent schizophrenia relapse prediction using mobile
  sensor based daily behavioral rhythm changes.
\newblock In {\em International Conference on Wireless Mobile Communication and
  Healthcare}, pages 18--33. Springer, 2020.

\bibitem{thati2022novel}
Ravi~Prasad Thati, Abhishek~Singh Dhadwal, Praveen Kumar, et~al.
\newblock A novel multi-modal depression detection approach based on mobile
  crowd sensing and task-based mechanisms.
\newblock {\em Multimedia Tools and Applications}, pages 1--34, 2022.

\bibitem{faurholt2021daily}
Maria Faurholt-Jepsen, Jonas Busk, Maj Vinberg, Ellen~Margrethe Christensen,
  Mads Frost, Jakob~E Bardram, Lars~Vedel Kessing, et~al.
\newblock Daily mobility patterns in patients with bipolar disorder and healthy
  individuals.
\newblock {\em Journal of Affective Disorders}, 278:413--422, 2021.

\bibitem{Zalzala2011}
A.~Zalzala, S.~Chia, L.~Zalzala, and A.~Karimi.
\newblock Healthcare technologies in developing countries.
\newblock In {\em 2011 IEEE GCC Conference and Exhibition (GCC)}, pages
  629--632, 2011.

\bibitem{Naydenova2015}
E.~Naydenova, A.~Tsanas, C.~Casals-Pascual, and M.~De Vos.
\newblock Smart diagnostic algorithms for automated detection of childhood
  pneumonia in resource-constrained settings.
\newblock In {\em 2015 IEEE Global Humanitarian Technology Conference (GHTC)},
  pages 377--384, 2015.

\bibitem{Shamiluulu2017}
S.~Shamiluulu, M.~M. Boukar, and Z.~Yussupova.
\newblock Medical tool for assisting patients in kazakhstan polyclinics.
\newblock In {\em 2017 13th International Conference on Electronics, Computer
  and Computation (ICECCO)}, pages 1--5, 2017.

\bibitem{Mitra2008}
S.~Mitra, M.~Mitra, and B.~B. Chaudhuri.
\newblock Rural cardiac healthcare system-a scheme for developing countries.
\newblock In {\em TENCON 2008 - 2008 IEEE Region 10 Conference}, pages 1--5,
  2008.

\bibitem{Ring2004}
B.~Ring.
\newblock Healthcare infrastructure for developing countries.
\newblock In {\em IEE Seminar on Appropriate Medical Technology for Developing
  Countries (Ref. No. 2000/014)}, pages 18/1--18/4, 2004.

\bibitem{Nabiev2008}
R.~Nabiev, S.~Eshonkhojaeva, N.~Rakhimi, R.~Blom, I.~Munabi, E.~Mukooyo, and
  L.~Smedman.
\newblock Framework for low-cost maintenance of computer infrastructure
  applicable in low and middle income countries.
\newblock In {\em The 12th IEEE International Conference on e-Health
  Networking, Applications and Services}, pages 143--147, 2008.

\bibitem{Cordero2014}
I.~{Cordero}.
\newblock Innovation for medical devices for low resource settings- how
  limiting the scope of work to creating low cost attention grabbing designs is
  not enough.
\newblock In {\em Appropriate Healthcare Technologies for Low Resource Settings
  (AHT 2014)}, pages 1--3, 2014.

\bibitem{Neighbour2014}
R.~Neighbour and R.~Eltringham.
\newblock The reality of designing appropriate 'low cost' medical products for
  developing countries and their unintended consequences.
\newblock In {\em Appropriate Healthcare Technologies for Low Resource Settings
  (AHT 2014)}, pages 1--4, 2014.

\bibitem{vasan2020}
Aditya Vasan and James Friend.
\newblock Medical devices for low-and middle-income countries: A review and
  directions for development.
\newblock {\em Journal of medical devices}, 14(1), 2020.

\bibitem{paradkar2017}
Neeraj Paradkar and Shubhajit~Roy Chowdhury.
\newblock Cardiac arrhythmia detection using photoplethysmography.
\newblock In {\em 2017 39th Annual International Conference of the IEEE
  Engineering in Medicine and Biology Society (EMBC)}, pages 113--116. IEEE,
  2017.

\bibitem{chang2022atrial}
Po-Cheng Chang, Ming-Shien Wen, Chung-Chuan Chou, Chun-Chieh Wang, and Kuo-Chun
  Hung.
\newblock Atrial fibrillation detection using ambulatory smartwatch
  photoplethysmography and validation with simultaneous holter recording.
\newblock {\em American Heart Journal}, 247:55--62, 2022.

\bibitem{antink2019}
Christoph~Hoog Antink, Simon Lyra, Michael Paul, Xinchi Yu, and Steffen
  Leonhardt.
\newblock A broader look: Camera-based vital sign estimation across the
  spectrum.
\newblock {\em Yearbook of medical informatics}, 28(1):102, 2019.

\bibitem{bae2022prospective}
Sean Bae, Silviu Borac, Yunus Emre, Jonathan Wang, Jiang Wu, Mehr Kashyap,
  Si-Hyuck Kang, Liwen Chen, Melissa Moran, Julie Cannon, et~al.
\newblock Prospective validation of smartphone-based heart rate and respiratory
  rate measurement algorithms.
\newblock {\em Communications Medicine}, 2(1):1--10, 2022.

\bibitem{WHO_ehealth}
Who | ehealth at who.
\newblock \url{http://www.emro.who.int/health-topics/ehealth/}.
\newblock Last accessed: 2022-05-14.

\bibitem{Iluyemi2008}
A.~Iluyemi, J.~Briggs, and R.~A. Burger.
\newblock Health service delivery in developing countries through ehealth:
  Making the case for low-cost wireless infrastructures.
\newblock In {\em 2008 5th IET Seminar on Appropriate Healthcare Technologies
  for Developing Countries}, pages 1--6, 2008.

\bibitem{Omary2009}
Z.~Omary, D.~Lupiana, F.~Mtenzi, and B.~Wu.
\newblock Challenges to e-healthcare adoption in developing countries: A case
  study of tanzania.
\newblock In {\em 2009 First International Conference on Networked Digital
  Technologies}, pages 201--209, 2009.

\bibitem{care2x.org}
Care2x.
\newblock \url{http://www.care2x.org/}.
\newblock Last accessed: 2022-05-14.

\bibitem{openehr}
{OpenEHR}.
\newblock \url{https://www.openehr.org/}.
\newblock Last accessed: 2022-05-14.

\bibitem{Shah2013}
Kamal Shah, Tara Slough, Ping Yeh, Suave Gombwa, Athanase Kiromera, Maria Oden,
  and Rebecca Kortum.
\newblock Novel open-source electronic medical records system for palliative
  care in low-resource settings.
\newblock {\em BMC palliative care}, 12:31, August 2013.

\bibitem{Lakshmi2011}
M.~D. Lakshmi and J.~P.~M. Dhas.
\newblock An open source private cloud solution for rural healthcare.
\newblock In {\em 2011 International Conference on Signal Processing,
  Communication, Computing and Networking Technologies}, pages 670--674, 2011.

\bibitem{Torre2013}
I.~de~la Torre, B.~Martínez, and M.~López-Coronado.
\newblock Analyzing open-source and commercial ehr solutions from an
  international perspective.
\newblock In {\em 2013 IEEE 15th International Conference on e-Health
  Networking, Applications and Services (Healthcom 2013)}, pages 399--403,
  2013.

\bibitem{hosxp}
{HOSxP}.
\newblock \url{https://sourceforge.net/projects/hosxp/}.
\newblock Last accessed: 2022-05-14.

\bibitem{openemr}
{OpenEMR}.
\newblock \url{https://www.open-emr.org/}.
\newblock Last accessed: 2022-05-14.

\bibitem{openvista}
{OpenVista}.
\newblock \url{https://sourceforge.net/projects/openvista/}.
\newblock Last accessed: 2022-05-14.

\bibitem{berrueta2021maternal}
Mabel Berrueta, Agustin Ciapponi, Ariel Bardach, Federico~Rodriguez Cairoli,
  Fabricio~J Castellano, Xu~Xiong, Andy Stergachis, Sabra Zaraa, Ajoke
  Sobanjo-ter Meulen, and Pierre Buekens.
\newblock Maternal and neonatal data collection systems in low-and
  middle-income countries for maternal vaccines active safety surveillance
  systems: A scoping review.
\newblock {\em BMC pregnancy and childbirth}, 21(1):1--19, 2021.

\bibitem{purkayastha2019comparison}
Saptarshi Purkayastha, Roshini Allam, Pallavi Maity, and Judy~W Gichoya.
\newblock Comparison of open-source electronic health record systems based on
  functional and user performance criteria.
\newblock {\em Healthcare informatics research}, 25(2):89--98, 2019.

\bibitem{Goel2017}
N.~A. Goel, A.~A. Alam, E.~M.~R. Eggert, and S.~Acharya.
\newblock Design and development of a customizable telemedicine platform for
  improving access to healthcare for underserved populations.
\newblock In {\em 2017 39th Annual International Conference of the IEEE
  Engineering in Medicine and Biology Society (EMBC)}, pages 2658--2661, 2017.

\bibitem{Sharwardy2013}
S.~N. Sharwardy, Z.~Rahman, S.~Parveen, H.~Sarwar, and A.~M. Hossain.
\newblock A cost-effective web-based teleconsultation system.
\newblock In {\em 2013 8th International Conference on Information Technology
  in Asia (CITA)}, pages 1--4, 2013.

\bibitem{Pambudi2004}
I.~T. Pambudi, T.~Hayasaka, T.~Tsubota, S.~Wada, and T.~Yamaguchi.
\newblock Sustainable patient information network (spain) for primary care
  health center in indonesia.
\newblock In {\em Proceedings of the 25th Annual International Conference of
  the IEEE Engineering in Medicine and Biology Society (IEEE Cat.
  No.03CH37439)}, volume~2, pages 1421--1424 Vol.2, 2004.

\bibitem{Chen2011}
W.~Chen and M.~Akay.
\newblock Developing emrs in developing countries.
\newblock {\em IEEE Transactions on Information Technology in Biomedicine},
  15(1):62--65, 2011.

\bibitem{filemaker}
Filemaker.
\newblock \url{https://www.claris.com/filemaker/}.
\newblock Last accessed: 2022-05-14.

\bibitem{Manya2018}
A.~Manya, S.~Sahay, J.~Braa, and B.~Shisia.
\newblock Understanding the effects of decentralization on health information
  systems in developing countries: A case of devolution in kenya.
\newblock In {\em 2018 IST-Africa Week Conference (IST-Africa)}, pages Page 1
  of 11--Page 11 of 11, 2018.

\bibitem{Nunziata2002}
E.~Nunziata, M.~Sumalgy, P.~Chongo, and H.~Sitoi.
\newblock Healthcare technology information system: the case study of
  mozambique with an eye on global approach for developing and in-transition
  countries.
\newblock In {\em IEE Seminar on Appropriate Medical Technology for Developing
  Countries (Ref. No. 2002/057)}, pages 2/1--2/11, 2002.

\bibitem{Martinez2005}
Andrés Martínez, Valentín Villarroel~Ortega, Joaquín Pascual, and Francisco
  Del Pozo~Guerrero.
\newblock Analysis of information and communication needs in rural primary
  healthcare in developing countries.
\newblock {\em IEEE transactions on information technology in biomedicine : a
  publication of the IEEE Engineering in Medicine and Biology Society},
  9:66--72, April 2005.

\bibitem{Chaamwe2010}
N.~Chaamwe.
\newblock Telehealth services in the context of zambia, a developing country.
\newblock In {\em 2010 Second International Conference on Information
  Technology and Computer Science}, pages 470--473, 2010.

\bibitem{Grainger2006}
P.~Grainger, D.~Kane, D.~Bowles, J.~Mahady, P.~Pentony, J.~G. Mahady,
  P.~Grainger, D.~Kane, D.~Bowles, J.~Mahady, P.~Pentony, and J.~G. Mahady.
\newblock Wireless communication and its application in the healthcare systems
  of developing countries.
\newblock In {\em 2006 The 4th Institution of Engineering and Technology
  Seminar on Appropriate Healthcare Technologies for Developing Countries},
  pages 97--104, 2006.

\bibitem{Martinez2002}
A.~Martinez, V.~Villarroel, J.~Seoane, and F.~del Pozo.
\newblock Ehas program: rural telemedicine systems for primary healthcare in
  developing countries.
\newblock In {\em IEEE 2002 International Symposium on Technology and Society
  (ISTAS'02). Social Implications of Information and Communication Technology.
  Proceedings (Cat. No.02CH37293)}, pages 31--36, 2002.

\bibitem{magaia2021artificial}
Naercio Magaia, Igor de~L Ribeiro, Andr{\'e}~WO de~Aguiar, Ramon Fonseca, Khan
  Muhammad, and Victor Hugo~C de~Albuquerque.
\newblock An artificial intelligence application for drone-assisted 5g remote
  e-health.
\newblock {\em IEEE Internet of Things Magazine}, 4(4):30--35, 2021.

\bibitem{shaylor2021tale}
Ruth~M Shaylor, Ian~T Chao, Jeremy~S Young, and Jasamine~M Coles-Black.
\newblock A tale of three countries: Three-dimensional printing for procedural
  simulation in the digital era.
\newblock {\em Anaesthesia and intensive care}, 49(2):140--143, 2021.

\bibitem{Puustjaervi2011}
J.~Puustjärvi and L.~Puustjärvi.
\newblock Automating remote monitoring and information therapy: An opportunity
  to practice telemedicine in developing countries.
\newblock In {\em 2011 IST-Africa Conference Proceedings}, pages 1--9, 2011.

\bibitem{Emmerling2014}
D.~A. Emmerling, R.~Sridhara, and R.~A. Malkin.
\newblock An open-source bmet library: results on access and value.
\newblock In {\em Appropriate Healthcare Technologies for Low Resource Settings
  (AHT 2014)}, pages 1--3, 2014.

\bibitem{Ruijter2008}
P.~de~Ruijter, G.~Ferreira, and R.~Parsons.
\newblock Using educational technology to reach a wider audience for healthcare
  technology management.
\newblock In {\em 2008 5th IET Seminar on Appropriate Healthcare Technologies
  for Developing Countries}, pages 1--4, 2008.

\bibitem{Bondale2013}
N.~Bondale, S.~Kimbahune, and A.~K. Pande.
\newblock mhealthphc: An ict tool for primary healthcare in india.
\newblock {\em IEEE Technology and Society Magazine}, 32(3):31--38, 2013.

\bibitem{opennmt}
Guillaume Klein, Yoon Kim, Yuntian Deng, Jean Senellart, and Alexander Rush.
\newblock {O}pen{NMT}: Open-source toolkit for neural machine translation.
\newblock In {\em Proceedings of {ACL} 2017, System Demonstrations}, pages
  67--72, Vancouver, Canada, July 2017. Association for Computational
  Linguistics.

\bibitem{Amararachchi2013}
J.~L. Amararachchi, H.~S.~C. Perera, and K.~Pulasinghe.
\newblock Knowledge management framework for achieving quality of healthcare in
  the developing countries.
\newblock In {\em 2013 International Conference on Computer Medical
  Applications (ICCMA)}, pages 1--6, 2013.

\bibitem{ali2019}
Abdullrahim Ali.
\newblock {\em e-Health systems adoption and telemedicine readiness:
  practitioner perspective from Libyan healthcare sector}.
\newblock PhD thesis, Brunel University London, 2019.

\bibitem{Montalban2008}
J.~M. Montalban and A.~B. Marcelo.
\newblock Information and communications technology needs assessment of
  philippine rural health physicians.
\newblock In {\em HealthCom 2008 - 10th International Conference on e-health
  Networking, Applications and Services}, pages 130--133, 2008.

\bibitem{Brown2014}
S.~Brown and K.~Rudahinduka.
\newblock Use of mobile devices for medical services in resource-limited
  settings: case study in rwanda.
\newblock In {\em Appropriate Healthcare Technologies for Low Resource Settings
  (AHT 2014)}, pages 1--4, 2014.

\bibitem{Hebert2016}
E.~Hebert, W.~Ferguson, S.~McCullough, M.~Chan, A.~Drobakha, S.~Ritter, and
  K.~Mehta.
\newblock mbody health: Digitizing disabilities in sierra leone.
\newblock In {\em 2016 IEEE Global Humanitarian Technology Conference (GHTC)},
  pages 717--724, 2016.

\bibitem{Mukherjee2012}
C.~Mukherjee, K.~Gupta, and R.~Nallusamy.
\newblock A system to provide primary maternity healthcare services in
  developing countries.
\newblock In {\em Annual SRII Global Conference}, pages 243--249, 2012.

\bibitem{Cardiax}
{Cardiax Mobile ECG}.
\newblock \url{https://www.cardiax.eu/t-en/mobil-app}.
\newblock Last accessed: 2022-05-14.

\bibitem{runtastic}
{Runtastic Heart Rate}.
\newblock \url{https://www.runtastic.com/}.
\newblock Last accessed: 2022-05-14.

\bibitem{HeartRate}
{Heart Rate Monitor}.
\newblock
  \url{https://play.google.com/store/apps/details?id=com.repsi.heartrate}.
\newblock Last accessed: 2022-05-14.

\bibitem{BloodPressureWatch}
{Blood Pressure Watch}.
\newblock
  \url{https://play.google.com/store/apps/details?id=com.boxeelab.healthlete.bpwatch}.
\newblock Last accessed: 2022-05-14.

\bibitem{FingerBloodPressure}
{Instant Blood Pressure}.
\newblock \url{https://www.instantbloodpressure.com/}.
\newblock Last accessed: 2022-05-14.

\bibitem{FingerPrintThermometer}
{Finger Print Thermometer}.
\newblock
  \url{https://fingerprint-thermometer.en.uptodown.com/android/download}.
\newblock Last accessed: 2022-05-14.

\bibitem{BodyTemperature}
Body temperature fever thermometer diary.
\newblock
  \url{https://play.google.com/store/apps/details?id=com.interactivespecializedsoftware.bodytemperature}.
\newblock Last accessed: 2022-05-14.

\bibitem{ioximeter}
{iOximeter}.
\newblock \url{http://safeheartus.com/ioximeter}.
\newblock Last accessed: 2022-05-14.

\bibitem{eyecareplus}
{Eye care plus}.
\newblock \url{https://eye-care-plus.en.uptodown.com/android}.
\newblock Last accessed: 2022-05-14.

\bibitem{testyourhearing}
{Test Your Hearing}.
\newblock \url{https://www.mimi.io}.
\newblock Last accessed: 2022-05-14.

\bibitem{uhear}
{uHear}.
\newblock \url{https://apps.apple.com/us/app/uhear/id309811822}.
\newblock Last accessed: 2022-05-14.

\bibitem{healtassistant}
{Health Assistant}.
\newblock \url{https://play.google.com/store/apps/details?id=com.wsmrs.hassi}.
\newblock Last accessed: 2022-05-14.

\bibitem{asthmatracker}
{Asthma Tracker: PEFLog}.
\newblock \url{https://www.tekemo.fi/peflog}.
\newblock Last accessed: 2022-05-14.

\bibitem{diabetesdiagnostics}
{Diabetes Diagnostics}.
\newblock
  \url{https://play.google.com/store/apps/details?id=com.szyk.diabetes}.
\newblock Last accessed: 2022-05-14.

\bibitem{wateryourbody}
{Water your body}.
\newblock \url{http://www.foware.com/}.
\newblock Last accessed: 2022-05-14.

\bibitem{medisafe}
{Medisafe Meds \& Pill Reminder}.
\newblock \url{https://www.medisafeapp.com/}.
\newblock Last accessed: 2022-05-14.

\bibitem{dosecast}
{Dosecast Indication Reminder}.
\newblock \url{http://www.montunosoftware.com/}.
\newblock Last accessed: 2022-05-14.

\bibitem{myfitnesspal}
My fitness pal.
\newblock \url{https://www.myfitnesspal.com/}.
\newblock Last accessed: 2022-05-14.

\bibitem{sleepcycle}
{Sleep Cycle}.
\newblock \url{https://www.sleepcycle.com/}.
\newblock Last accessed: 2022-05-14.

\bibitem{fitbit}
Fitbit.
\newblock \url{https://fitbit.com/}.
\newblock Last accessed: 2022-05-14.

\bibitem{rubin2015characterizing}
Geoffrey~D Rubin, Justus~E Roos, Martin Tall, Brian Harrawood, Sukantadev Bag,
  Donald~L Ly, Danielle~M Seaman, Lynne~M Hurwitz, Sandy Napel, and Kingshuk
  Roy~Choudhury.
\newblock Characterizing search, recognition, and decision in the detection of
  lung nodules on ct scans: elucidation with eye tracking.
\newblock {\em Radiology}, 274(1):276--286, 2015.

\bibitem{Hosny2018}
Ahmed Hosny, Chintan Parmar, John Quackenbush, Lawrence~H. Schwartz, and Hugo
  J. W.~L. Aerts.
\newblock Artificial intelligence in radiology.
\newblock {\em Nature reviews. Cancer}, 18(29777175):500--510, August 2018.

\bibitem{Avati2018}
Anand Avati, Stephen Pfohl, Chris Lin, Thao Nguyen, Meng Zhang, Philip Hwang,
  Jessica Wetstone, Kenneth Jung, Andrew Ng, and Nigam Shah.
\newblock {\em Predicting Inpatient Discharge Prioritization With Electronic
  Health Records}.
\newblock December 2018.

\bibitem{ahn2021machine}
Imjin Ahn, Hansle Gwon, Heejun Kang, Yunha Kim, Hyeram Seo, Heejung Choi, Ha~Na
  Cho, Minkyoung Kim, Tae~Joon Jun, Young-Hak Kim, et~al.
\newblock Machine learning--based hospital discharge prediction for patients
  with cardiovascular diseases: Development and usability study.
\newblock {\em JMIR Medical Informatics}, 9(11):e32662, 2021.

\bibitem{chen2022machine}
Talen Chen, Samaneh Madanian, David Airehrour, and Marianne Cherrington.
\newblock Machine learning methods for hospital readmission prediction:
  systematic analysis of literature.
\newblock {\em Journal of Reliable Intelligent Environments}, pages 1--18,
  2022.

\bibitem{Hannun2019}
Awni~Y. Hannun, Pranav Rajpurkar, Masoumeh Haghpanahi, Geoffrey~H. Tison, Codie
  Bourn, Mintu~P. Turakhia, and Andrew~Y. Ng.
\newblock Cardiologist-level arrhythmia detection and classification in
  ambulatory electrocardiograms using a deep neural network.
\newblock {\em Nature Medicine}, 25(1):65--69, January 2019.

\bibitem{jo2021detection}
Yong-Yeon Jo, Joon-myoung Kwon, Ki-Hyun Jeon, Yong-Hyeon Cho, Jae-Hyun Shin,
  Yoon-Ji Lee, Min-Seung Jung, Jang-Hyeon Ban, Kyung-Hee Kim, Soo~Youn Lee,
  et~al.
\newblock Detection and classification of arrhythmia using an explainable deep
  learning model.
\newblock {\em Journal of Electrocardiology}, 67:124--132, 2021.

\bibitem{ccinar2021classification}
Ahmet {\c{C}}{\i}nar and Seda~Arslan Tuncer.
\newblock Classification of normal sinus rhythm, abnormal arrhythmia and
  congestive heart failure ecg signals using lstm and hybrid cnn-svm deep
  neural networks.
\newblock {\em Computer methods in biomechanics and biomedical engineering},
  24(2):203--214, 2021.

\bibitem{Esteva2017}
Andre Esteva, Brett Kuprel, Roberto Novoa, Justin Ko, Susan Swetter, Helen
  Blau, and Sebastian Thrun.
\newblock Dermatologist-level classification of skin cancer with deep neural
  networks.
\newblock {\em Nature}, 542, January 2017.

\bibitem{Voisin2018}
Maxime Voisin, Yichen Shen, Alireza Aliamiri, Anand Avati, Awni Hannun, and
  Andrew Ng.
\newblock {\em Ambulatory Atrial Fibrillation Monitoring Using Wearable
  Photoplethysmography with Deep Learning}.
\newblock November 2018.

\bibitem{jiang2017artificial}
Fei Jiang, Yong Jiang, Hui Zhi, Yi~Dong, Hao Li, Sufeng Ma, Yilong Wang, Qiang
  Dong, Haipeng Shen, and Yongjun Wang.
\newblock Artificial intelligence in healthcare: past, present and future.
\newblock {\em Stroke and vascular neurology}, 2(4), 2017.

\bibitem{yu2018artificial}
Kun-Hsing Yu, Andrew~L Beam, and Isaac~S Kohane.
\newblock Artificial intelligence in healthcare.
\newblock {\em Nature biomedical engineering}, 2(10):719--731, 2018.

\bibitem{reddy2019artificial}
Sandeep Reddy, John Fox, and Maulik~P Purohit.
\newblock Artificial intelligence-enabled healthcare delivery.
\newblock {\em Journal of the Royal Society of Medicine}, 112(1):22--28, 2019.

\bibitem{vaananen2021ai}
Antti V{\"a}{\"a}n{\"a}nen, Keijo Haataja, Katri Vehvil{\"a}inen-Julkunen, and
  Pekka Toivanen.
\newblock Ai in healthcare: A narrative review.
\newblock {\em F1000Research}, 10(6):6, 2021.

\bibitem{shaheen2021applications}
Mohammed~Yousef Shaheen.
\newblock Applications of artificial intelligence (ai) in healthcare: A review.
\newblock {\em ScienceOpen Preprints}, 2021.

\bibitem{Doshi2017}
Riddhi Doshi, Dennis Falzon, Bruce~V. Thomas, Zelalem Temesgen, Lal Sadasivan,
  Giovanni~Battista Migliori, and Mario Raviglione.
\newblock Tuberculosis control, and the where and why of artificial
  intelligence.
\newblock {\em ERJ open research}, 3(28656130):00056--2017, June 2017.

\bibitem{Brunskill2010}
Emma Brunskill and Neal Lesh.
\newblock {\em Routing for Rural Health: Optimizing Community Health Worker
  Visit Schedules.}
\newblock January 2010.

\bibitem{mesko2018will}
Bertalan Mesk{\'o}, Gergely Het{\'e}nyi, and Zsuzsanna Gy{\H{o}}rffy.
\newblock Will artificial intelligence solve the human resource crisis in
  healthcare?
\newblock {\em BMC health services research}, 18(1):1--4, 2018.

\bibitem{Krishnan2016}
B.~Krishnan, S.~Babu, S.~P. Shaji, A.~S.~R. Tamanampudi, and S.~S.~S.
  Sanagapati.
\newblock Software based gateway with distributed flow environment for medical
  iot in rural areas.
\newblock In {\em 2016 IEEE International Conference on Advanced Networks and
  Telecommunications Systems (ANTS)}, pages 1--5, 2016.

\bibitem{Ganesh2016}
G.~R.~D. Ganesh, K.~Jaidurgamohan, V.~Srinu, C.~R. Kancharla, and S.~V.~S.
  Suresh.
\newblock Design of a low cost smart chair for telemedicine and iot based
  health monitoring: An open source technology to facilitate better healthcare.
\newblock In {\em 2016 11th International Conference on Industrial and
  Information Systems (ICIIS)}, pages 89--94, 2016.

\bibitem{Raj2017}
C.~Raj, C.~Jain, and W.~Arif.
\newblock Heman: Health monitoring and nous: An iot based e-health care system
  for remote telemedicine.
\newblock In {\em 2017 International Conference on Wireless Communications,
  Signal Processing and Networking (WiSPNET)}, pages 2115--2119, 2017.

\bibitem{Divakaran2017}
S.~Divakaran, L.~Manukonda, N.~Sravya, M.~M. Morais, and P.~Janani.
\newblock Iot clinic-internet based patient monitoring and diagnosis system.
\newblock In {\em 2017 IEEE International Conference on Power, Control, Signals
  and Instrumentation Engineering (ICPCSI)}, pages 2858--2862, 2017.

\bibitem{Prakashan2017}
K.~Prakashan, A.~S. Karthika, R.~Ankayarkanni, and J.~B. Jose.
\newblock Transformation of health care system using internet of things in
  villages.
\newblock In {\em 2017 IEEE International Conference on Industrial Engineering
  and Engineering Management (IEEM)}, pages 914--918, 2017.

\bibitem{Zilani2018}
K.~A. Zilani, R.~Yeasmin, K.~A. Zubair, M.~R. Sammir, and S.~Sabrin.
\newblock R3hms, an iot based approach for patient health monitoring.
\newblock In {\em 2018 International Conference on Computer, Communication,
  Chemical, Material and Electronic Engineering (IC4ME2)}, pages 1--4, 2018.

\bibitem{ryu2021comprehensive}
Dennis Ryu, Dong~Hyun Kim, Joan~T Price, Jong~Yoon Lee, Ha~Uk Chung, Emily
  Allen, Jessica~R Walter, Hyoyoung Jeong, Jingyue Cao, Elena Kulikova, et~al.
\newblock Comprehensive pregnancy monitoring with a network of wireless, soft,
  and flexible sensors in high-and low-resource health settings.
\newblock {\em Proceedings of the National Academy of Sciences}, 118(20), 2021.

\bibitem{sriraam2021wearable}
Natarajan Sriraam, Prabhu~Ravikala Vittal, Uma Arun, Karthik Angadi, Arjun
  Halli, and Priyanka Chakravarthy.
\newblock Wearable wireless multiparameter monitoring systems (wemums) for
  resource constrained settings.
\newblock In {\em 2021 3rd International Conference on Electrical, Control and
  Instrumentation Engineering (ICECIE)}, pages 1--4. IEEE, 2021.

\bibitem{dese2022low}
Kokeb Dese, Gelan Ayana, and Gizeaddis~Lamesgin Simegn.
\newblock Low cost, non-invasive, and continuous vital signs monitoring device
  for pregnant women in low resource settings (lvital device).
\newblock {\em HardwareX}, 11:e00276, 2022.

\bibitem{Sarkar2017}
S.~Sarkar and R.~Saha.
\newblock A futuristic iot based approach for providing healthcare support
  through e-diagnostic system in india.
\newblock In {\em 2017 Second International Conference on Electrical, Computer
  and Communication Technologies (ICECCT)}, pages 1--7, 2017.

\bibitem{Pathinarupothi2018}
R.~K. Pathinarupothi, P.~Durga, and E.~S. Rangan.
\newblock Iot-based smart edge for global health: Remote monitoring with
  severity detection and alerts transmission.
\newblock {\em IEEE Internet of Things Journal}, 6(2):2449--2462, 2018.

\bibitem{Islam2015}
S.~M.~R. Islam, D.~Kwak, M.~H. Kabir, M.~Hossain, and K.~Kwak.
\newblock The internet of things for health care: A comprehensive survey.
\newblock {\em IEEE Access}, 3:678--708, 2015.

\bibitem{weiss2017blockchain}
Martin Weiss, Ad{\`e}le Botha, Marlien Herselman, and Glaudina Loots.
\newblock Blockchain as an enabler for public mhealth solutions in south
  africa.
\newblock In {\em 2017 IST-Africa Week Conference (IST-Africa)}, pages 1--8.
  IEEE, 2017.

\bibitem{kamau2018blockchain}
Gabriel Kamau, Caroline Boore, Elizaphan Maina, and Stephen Njenga.
\newblock Blockchain technology: Is this the solution to emr interoperability
  and security issues in developing countries?
\newblock In {\em 2018 IST-Africa Week Conference (IST-Africa)}, pages Page--1.
  IEEE, 2018.

\bibitem{ayubblockchain}
Rana~Mubashar Ayub and Nadeem Javaid.
\newblock A blockchain-based incentive mechanism for patient-driven data
  collection in the healthcare.

\bibitem{theodouli2018design}
Anastasia Theodouli, Stelios Arakliotis, Konstantinos Moschou, Konstantinos
  Votis, and Dimitrios Tzovaras.
\newblock On the design of a blockchain-based system to facilitate healthcare
  data sharing.
\newblock In {\em 2018 17th IEEE International Conference On Trust, Security
  And Privacy In Computing And Communications/12th IEEE International
  Conference On Big Data Science And Engineering (TrustCom/BigDataSE)}, pages
  1374--1379. IEEE, 2018.

\bibitem{liang2017integrating}
Xueping Liang, Juan Zhao, Sachin Shetty, Jihong Liu, and Danyi Li.
\newblock Integrating blockchain for data sharing and collaboration in mobile
  healthcare applications.
\newblock In {\em 2017 IEEE 28th annual international symposium on personal,
  indoor, and mobile radio communications (PIMRC)}, pages 1--5. IEEE, 2017.

\bibitem{alexaki2018blockchain}
Sofia Alexaki, George Alexandris, Vasilis Katos, and Nikolaos~E Petroulakis.
\newblock Blockchain-based electronic patient records for regulated circular
  healthcare jurisdictions.
\newblock In {\em 2018 IEEE 23rd International Workshop on Computer Aided
  Modeling and Design of Communication Links and Networks (CAMAD)}, pages 1--6.
  IEEE, 2018.

\bibitem{medicalchain}
MedicalChain.
\newblock Whitepaper 2 - medicalchain.
\newblock \url{https://medicalchain.com/Medicalchain-Whitepaper-EN.pdf}.
\newblock Last accessed: 2022-05-14.

\bibitem{lam018applications}
Gajendra~J Katuwal, Sandip Pandey, Mark Hennessey, and Bishal Lamichhane.
\newblock Applications of blockchain in healthcare: current landscape \&
  challenges.
\newblock {\em arXiv preprint arXiv:1812.02776}, 2018.

\bibitem{holbl2018systematic}
Marko H{\"o}lbl, Marko Kompara, Aida Kami{\v{s}}ali{\'c}, and Lili
  Nemec~Zlatolas.
\newblock A systematic review of the use of blockchain in healthcare.
\newblock {\em Symmetry}, 10(10):470, 2018.

\bibitem{mcghin2019blockchain}
Thomas McGhin, Kim-Kwang~Raymond Choo, Charles~Zhechao Liu, and Debiao He.
\newblock Blockchain in healthcare applications: Research challenges and
  opportunities.
\newblock {\em Journal of Network and Computer Applications}, 135:62--75, 2019.

\bibitem{attaran2022blockchain}
Mohsen Attaran.
\newblock Blockchain technology in healthcare: Challenges and opportunities.
\newblock {\em International Journal of Healthcare Management}, 15(1):70--83,
  2022.

\bibitem{sookhak2021blockchain}
Mehdi Sookhak, Mohammad~Reza Jabbarpour, Nader~Sohrabi Safa, and F~Richard Yu.
\newblock Blockchain and smart contract for access control in healthcare: a
  survey, issues and challenges, and open issues.
\newblock {\em Journal of Network and Computer Applications}, 178:102950, 2021.

\bibitem{panta2020compliance}
Gopal Panta, Ann~K Richardson, Ian~C Shaw, and Patricia~A Coope.
\newblock Compliance of primary and secondary care public hospitals with
  standard practices for reprocessing and steam sterilization of reusable
  medical devices in nepal: findings from nation-wide multicenter clustered
  audits.
\newblock {\em BMC health services research}, 20(1):1--13, 2020.

\bibitem{thapa2022effect}
Rita Thapa, Alison Yih, Ashish Chauhan, Salomi Poudel, Sagar Singh, Suresh
  Shrestha, Suresh Tamang, Rishav Shrestha, and Ruma Rajbhandari.
\newblock Effect of deploying biomedical equipment technician on the
  functionality of medical equipment in the government hospitals of rural
  nepal.
\newblock {\em Human resources for health}, 20(1):1--8, 2022.

\bibitem{singh2021potential}
Gurfarmaan Singh, Robert Casson, and WengOnn Chan.
\newblock The potential impact of 5g telecommunication technology on
  ophthalmology.
\newblock {\em Eye}, 35(7):1859--1868, 2021.

\bibitem{kumar2022importance}
Saurabh Kumar, Ankush Sharma, and Priyanka Rishi.
\newblock Importance and uses of telemedicine in physiotherapeutic healthcare
  system: A scoping systemic review.
\newblock {\em Data Engineering for Smart Systems}, pages 411--422, 2022.

\bibitem{minervini2022teledentistry}
Giuseppe Minervini, Diana Russo, Alan~Scott Herford, Francesca Gorassini, Aida
  Meto, Cesare D’Amico, Gabriele Cervino, Marco Cicci{\`u}, and Luca
  Fiorillo.
\newblock Teledentistry in the management of patients with dental and
  temporomandibular disorders.
\newblock {\em BioMed Research International}, 2022, 2022.

\bibitem{el2022telehealth}
Dina~M El-Sherif, Mohamed Abouzid, Mohamed~Tarek Elzarif, Alhassan~Ali Ahmed,
  Ashwag Albakri, and Mohammed~M Alshehri.
\newblock Telehealth and artificial intelligence insights into healthcare
  during the covid-19 pandemic.
\newblock In {\em Healthcare}, volume~10, page 385. MDPI, 2022.

\bibitem{archer2021ehealth}
Norman Archer, Cynthia Lokker, Maryam Ghasemaghaei, Deborah DiLiberto, et~al.
\newblock ehealth implementation issues in low-resource countries: model,
  survey, and analysis of user experience.
\newblock {\em Journal of medical Internet research}, 23(6):e23715, 2021.

\bibitem{adu2019individuals}
Ernest~K Adu, Nelly Todorova, and Annette Mills.
\newblock Do individuals in developing countries care about personal health
  information privacy? an empirical investigation.
\newblock In {\em CONF-IRM}, page~16, 2019.

\bibitem{hasan2022camsense}
Zahid Hasan, Sreenivasan~Ramasamy Ramamurthy, and Nirmalya Roy.
\newblock Camsense: A camera-based contact-less heart activity monitoring.
\newblock {\em Smart Health}, 23:100240, 2022.

\bibitem{lamichhane2022econet}
Bishal Lamichhane, Nidal Moukaddam, Ankit~B Patel, and Ashutosh Sabharwal.
\newblock Econet: Estimating everyday conversational network from free-living
  audio for mental health applications.
\newblock {\em IEEE Pervasive Computing}, (01):1--9, 2022.

\bibitem{thomason2021blockchain}
Jane Thomason, Sonja Bernhardt, Tia Kansara, and Nichola Cooper.
\newblock Blockchain for universal health coverage.
\newblock In {\em Research Anthology on Blockchain Technology in Business,
  Healthcare, Education, and Government}, pages 488--502. IGI Global, 2021.

\end{thebibliography}

\end{document}